\documentclass[a4paper,11pt,articletitle = true]{article}
\usepackage{dcolumn}
\usepackage[dvips]{epsfig,graphics}
\usepackage{threeparttable}
\usepackage{indentfirst}
\usepackage{ifthen}
\usepackage{rotating}
\usepackage{color}
\usepackage{caption}
\usepackage{subfig}
\usepackage{amssymb}
\usepackage{amsmath}
\usepackage{achemso}
\usepackage{pdflscape}
\usepackage{color,soul} % added by DC
\usepackage{float} % added by DC

\begin{document}

\title{\bf From Cyclic Nanorings to Single-Walled Carbon Nanotubes: Disclosing the Evolution of their Electronic Structure with the Help of Theoretical Methods}

\author{A. P\'erez-Guardiola$^a$, R. Ortiz-Cano$^{a,b,c}$, M. E. Sandoval-Salinas$^{d,e}$, \\ J. Fern\'andez-Rossier$^{b,c}$, D. Casanova$^{e,f}$, \\ 
A. J. P\'erez-Jim\'enez$^a$, and J. C. Sancho-Garc\'{\i}a$^a$\footnote{E-mail: jc.sancho@ua.es} 
\\ \\
$^a$ Department of Physical Chemistry, \\
University of Alicante, \\ 
E-03080 Alicante, Spain \\ \\
$^b$ Department of Applied Physics, \\
University of Alicante, \\
E-03080 Alicante, Spain \\ \\
$^c$ QuantaLab, \\
International Iberian Nanotechnology Laboratory (INL), \\
4715-330 Braga, Portugal \\ \\
$^d$ Departament de Ci\'encia de Materials i Qu\'{\i}mica F\'{\i}sica, \\
Institut de Qu\'{\i}mica Te\`orica i Computacional (IQTCUB), \\
Universitat de Barcelona, \\
E-08028 Barcelona, Spain \\ \\
$^e$ Donostia International Physics Center (DIPC), \\
E-20018 Donostia, Spain \\ \\
$^f$ IKERBASQUE, \\
Basque Foundation for Science, \\
E-48013 Bilbao, Spain \\ \\
}

\date{\today}

\maketitle

\clearpage
\normalsize

\begin{abstract}
\setlength{\baselineskip}{0.35in}

We systematically investigate the relationships between structural and electronic effects of finite size zigzag or armchair carbon nanotubes of various diameters and lengths, 
starting from a molecular template of varying shape and diameter, i.e. cyclic oligoacene or oligophenacene molecules, and disclosing how adding layers and/or end-caps (i.e. 
hemifullerenes) can modify their (poly)radicaloid nature. We mostly used tight-binding and finite-temperature density-based methods, the former providing a simple but intuitive 
picture about their electronic structure, and the latter dealing effectively with strong correlation effects by relying on a fractional occupation number weighted electron density 
($\rho^{\mathrm{FOD}}$), with additional RAS-SF calculations backing up the latter results. We also explore how minor structural modifications of nanotube end-caps might influence 
the results, showing that topology, together with the chemical nature of the systems, is pivotal for the understanding of the electronic properties of these and other related 
systems.

\end{abstract}


\clearpage

% main text

\setlength{\baselineskip}{0.3in}

\section{Introduction}

Over the past decades, there has been a growing interest for studying open-shell electronic structures of Polycyclic Aromatic Hydrocarbons (PAH). These polyradicaloid molecules, 
often with planar conjugated backbones, were synthetically elusive until very recently when phenalenyl, or in general larger triangulene-like derivatives, were experimentally 
obtained with great promise for technological applications\cite{inoue2001,morita2011,das2016,wang2016}. All these molecules are traditionally categorized as open-shell 
non-Kekul\'e systems, containing unpaired electrons due to topological effects. However, further extension to polyradical backbones such as zethrenes\cite{lukman2017}, or in 
general to nanographene radicals \cite{desroches2017}, has recently shown particularly interesting singlet-fission\cite{lopez2017} or non-linear optics capabilities\cite{nakano2015}, 
to name just a few of envisioned applications to be exploited in photophysics or excitonic-based fields. The ground state spin multiplicity of alternating hydrocarbons can be 
predicted by means of the Ovchinnikov's rule, later generalized by the Lieb's theorem\cite{lieb1989}, for which the spin quantum number of a system is given by the expression 
$S = \frac{\vert N_A - N_B \vert}{2}$, with $N_A$ ($N_B$) the number of C atoms in a bipartite lattice. 
Consequently, the energetical (near-)degeneracy of their one-electron electronic states, also known as zero-energy modes, is given by the sublattice imbalance $N_Z = N_A - N_B$. 
On the other hand, there is a large variety of PAHs with a singlet ground-state but nevertheless a significant bi- or poly-radical character, with long acenes being probably the 
paradigmatic example. The latter systems are known to exhibit an increasing poly-radical character with the number of fused benzene rings, with systems longer than pentacene 
experimentally reported not so far ago and not without significant synthetic efforts\cite{mondal2006,tonshoff2010,huang2016,einholz2017}.
\\

\begin{figure}
\begin{center}
\scalebox{0.5}{\includegraphics{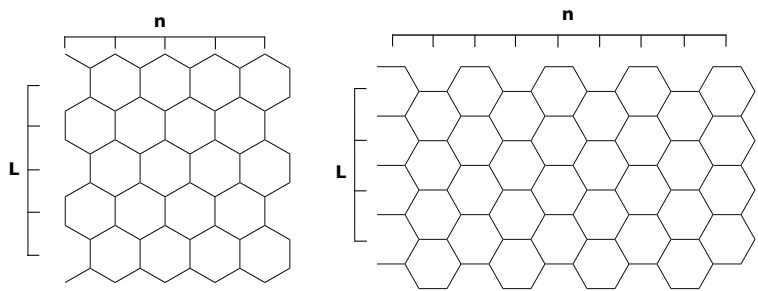}} \\
\caption{Chemical structures of [$n$]CC (left) and [$n$]CPH (right) systems, with H atoms omitted for clarity. The number of fused rings ($n$) and the length ($L$) of 
the corresponding nanotube are also indicated.
\label{fig:structures}}
\end{center}
\end{figure}

This well-established scenario has been altered after the recent emergence of cyclic nanorings and nanobelts, i.e. cyclic organic molecules of limited size and specific edges, 
constituted by fused and tilted benzene rings\cite{omachi2012,segawa2012,hirst2012,yamago2014,golder2015,kayahara2015,darzi2015,segawa2016} until closing the loop. Note that 
these molecules represent the shortest segment of armchair or zigzag Single-Walled Carbon NanoTubes (SWCNT) and are thus envisioned as molecular templates for the growth of 
SWCNT of controlled diameter and shape. Whereas CycloParaPhenylene (CPP) compounds are synthesized in a variety of experimental conditions and yields, CyclaCenes (CC) are very 
elusive so far\cite{lu2017} with their reactivity possibly related to their polyradical (di- and tetra- depending on the even or odd number of rings, $n$, respectively) character 
as it was previously argued\cite{choi1999,chen2007,sadowsky2010} and recently disclosed by more sophisticated state-of-the-art theoretical methods\cite{perez-guardiola2018}. As 
an intermediate case, we find CycloPHenacene (CPH) compounds, for which an 
isomer of [12]CPH ([3]cyclobenzo[a]anthracene) has been successfully synthesized very recently\cite{povie2017}. Figure~\ref{fig:structures} shows the chemical structure of 
[$n$]CC and [$n$]CPH, with $n$ indicating the number of fused rings forming the cyclic structures. However, how and why the electronic structure of these finite-size nanorings 
might change with their structural features (i.e. diameter -$n$-, form of the edges, length -$L$-, and caps) towards the step-by-step formation of SWCNT, is still a matter of 
investigation that we will systematically tackle here. 
\\

Therefore, taking necessarily into account the expected open-shell electronic structure of these cyclic systems, the choice of the theoretical method is far from being 
trivial. Multi-Configurational (MC) methods have been used in the past for open-shell systems with great success\cite{szalay2011}, but their pronounced computational scaling 
with the system size precludes its application to linear, planar, or cyclic PAHs composed of more than a few tens of carbon atoms. On the other hand, the cost-effective Density-Functional 
Theory (DFT) has been traditionally discarded for these (partly) open-shell systems because of the spin-contamination problem and/or the historical difficulties to deal with 
(near-)degeneracy effects\cite{cohen2011,sanfabian2013}, noting some recent efforts to model the electronic structure of open-shell singlet biradicals by time-dependent 
versions\cite{canola2018}. Hence, another way to afford these challenging effects is by relying on a fractional occupation formalism, mimicking thus the occupancy 
of the open-shell configurations and preventing spin contamination issues at a reasonable computational cost. The fractional occupation is obtained by imposing a Fermi-Dirac (i.e. 
a Fermi-smearing) distribution under the effect of a fictitious temperature to force that occupation, often dubbed this technique as Finite-Temperature (FT-)DFT\cite{grimme2013} 
or Thermally-Assisted-Occupation (TAO-)DFT\cite{chai2012,chai2014}. This formalism also provides qualitatively right density distributions, and it could thus be applied to systems 
of any chemical nature and structural topology, particularly interesting for PAHs\cite{yeh2016}, which we will exploit here for pristine or end-capped [$n$]CC and [$n$]CPH cyclic 
compounds of various diameters and lengths. We finally compare these results with those from the Restricted-Active-Space Spin-Flip (RAS-SF) method\cite{casanova2009}.
\\

\section{Theoretical framework}

\subsection{The FT-DFT method}

Fractional orbital occupation associated with (near-)degeneracy effects, which in turns arise from static or non-dynamical correlation effects, becomes intrinsically difficult 
to treat by any standard DFT methodology. These fractional occupation numbers ($0 \leq f_i \leq 1$) are known to affect the electronic density, built from the set of 
occupied orbitals $\left\lbrace \varphi_i \right\rbrace$ self-consistently obtained, through the following expression:
\begin{equation}
\label{eqn:1}
\rho({\mathbf r}) = \sum_i^{\infty} f_i \vert \varphi_i ({\mathbf r}) \vert^2,
\end{equation}
and are determined from a Finite-Temperature (FT) Fermi-Dirac distribution\cite{lin2017}:
\begin{equation}
\label{eqn:2}
f_i = \frac{1}{1 + e^{(\epsilon_i - E_F)/\theta}},
\end{equation}
depending critically on the $\epsilon_i - E_F$ difference, with $\epsilon_i$ the eigenenergies of $\varphi_i$ and $E_F$ the Fermi level, and on $\theta = k_B T_{el}$, with 
$k_B$ the Boltzmann constant and $T_{el}$ a pseudo-temperature used to self-consistently minimize the Gibbs electronic free energy ($G_{el} = E_{el} - T_{el} S_{el}$) of 
the system. Once a set of fractionally occupied orbitals is generated following the above procedure, we can define a Fractional Occupation Density (FOD) as a real-space measure of 
static correlation effects\cite{grimme2015}:
\begin{equation}
\label{eqn:3}
\rho^{\mathrm{FOD}}({\mathbf r}) = \sum_i \left( \delta_1 - \delta_2 f_i \right) \vert \varphi_i ({\mathbf r}) \vert^2,
\end{equation}
with $\delta_1$ and $\delta_2$ chosen to be $(1,1)$ if the eigenenergy ($\epsilon_i$) is lower than the energy of the Fermi level, $E_F$, or $(0,-1)$ otherwise. This orbital-based 
representation has shown to display useful information about the distribution of unpaired electrons in molecular systems\cite{plasser2013,horn2014,das2016} and can be integrated:
\begin{equation}
\label{eqn:4a}
\mbox{N$_{\mathrm{FOD}}$} = \int \rho^{\mathrm{FOD}}({\mathbf r}) d{\mathbf r},
\end{equation}
to yield the N$_{\mathrm{FOD}}$ value estimating the number of strongly correlated electrons of the system. This cost-effective FT-DFT methodology has been applied before to 
a variety of chemical (bio)systems\cite{bauer2017} and to small-size linear and cyclic oligoacenes\cite{perez-guardiola2018}, with the latter results in good agreement with 
findings from the RAS-SF method\cite{casanova2009}, showing thus its potential to deal with larger systems out of the scope of more costly methods.
\\

\subsubsection{Computational details}

The ORCA 4.0.0.2 package\cite{neese2012} was used for all the FT-DFT calculations reported here. All compounds studied were optimized at the TPSS-D3(BJ)/def2-SVP level, 
i.e. employing a dispersion-corrected functional to efficiently incorporate intra-molecular non-covalent (i.e. dispersion) effects\cite{grimme2010}. The corresponding FOD-based 
calculations were done at the (unrestricted) TPSS/def2-TZVP level\cite{tao2003}, and with the default temperature $T_{el} = 5000$K as recommended for this 
functional\cite{grimme2015,bauer2017}. The use of a hybrid version (i.e. TPSS0 with a 25~\% of exact exchange) with a modified temperature (10000K according to the relationship 
$T_{el}/K = 5000 + 20000\cdot a_x$) was also tested without significantly 
modifying the results. Singlet, triplet, and charged energies were obtained by imposing the adequate charge and multiplicity, but without modifying the level of theory fixed 
above. We increased systematically in all cases the thresholds for SCF calculations (i.e. TightSCF) and numerical integration (i.e. Grid6, NoFinalGrid). The isocontour 
values for displaying the $\rho^{\mathrm{FOD}}({\mathbf r})$ density were consistently set to $0\mathrm{.}005$ e/bohr$^3$, and the plots were generated with the UCSF 
Chimera\cite{pettersen2004} (version 1.12) package after proper manipulation of output files. The N$_{\mathrm{FOD}}$ values are directly extracted from the output file of 
the calculations.
\\

\subsection{The RAS-SF method}

Spin-Flip (SF) methods rely on a excitation operator promoting $\alpha$ electrons into empty $\beta$ orbitals, which together with an adequate choice of an active space 
of orbitals, the Restricted Active Space (RAS), can deal with degeneracies or near-degeneracies of electronic states\cite{krylov2001,krylov2006}. Actually, the combination 
known as RAS-SF has allowed to accurately treat molecular systems with radical or polyradical character. The number of unpaired electrons is quantified through the 
expression\cite{headgordon2003}:
\begin{equation}
\label{eqn:4b}
\mbox{N$_{\mathrm{U}}$} = \sum_i \left( 1 - \mbox{abs} \left( 1 - n_i \right) \right),
\end{equation}
where $n_i$ are the electron occupancies of the RAS-SF natural orbitals $(0 \leq n_i \leq 2)$. Note that Eqs. (\ref{eqn:4a}) and (\ref{eqn:4b}) are equivalent to each other 
(i.e. $\mbox{N$_{\mathrm{FOD}}$} \equiv \mbox{N$_{\mathrm{U}}$}$) and actually the RAS-SF calculated fractionally occupied orbitals can be also used to define the corresponding fractional 
occupation density.
\\

\subsubsection{Computational details}

The RAS-SF calculations were performed with the Q-Chem program\cite{shao2015} using the lowest ROHF (Restricted Open-Shell) triplet state as a 
reference configuration in all the systems. The (restricted) active space consisted on all virtuals and doubly occupied $\pi$-orbitals for the RAS1 and RAS3 subspaces, 
respectively, with 8 electrons in 8 $\pi$-orbitals for the RAS2 subspace. We disregard excitations from core electrons (1s C orbitals) and to virtual orbitals with energies 
higher than 0.5 a.u. for computational efficiency, and use the cost-effective 6-31G(d) basis set herein.
\\

\section{Results and discussion}

\subsection{Physics provided by the tight-binding method}

\begin{landscape}
\begin{figure}
\begin{center}
\begin{tabular}{cccc}
(a) \scalebox{0.25}{\includegraphics{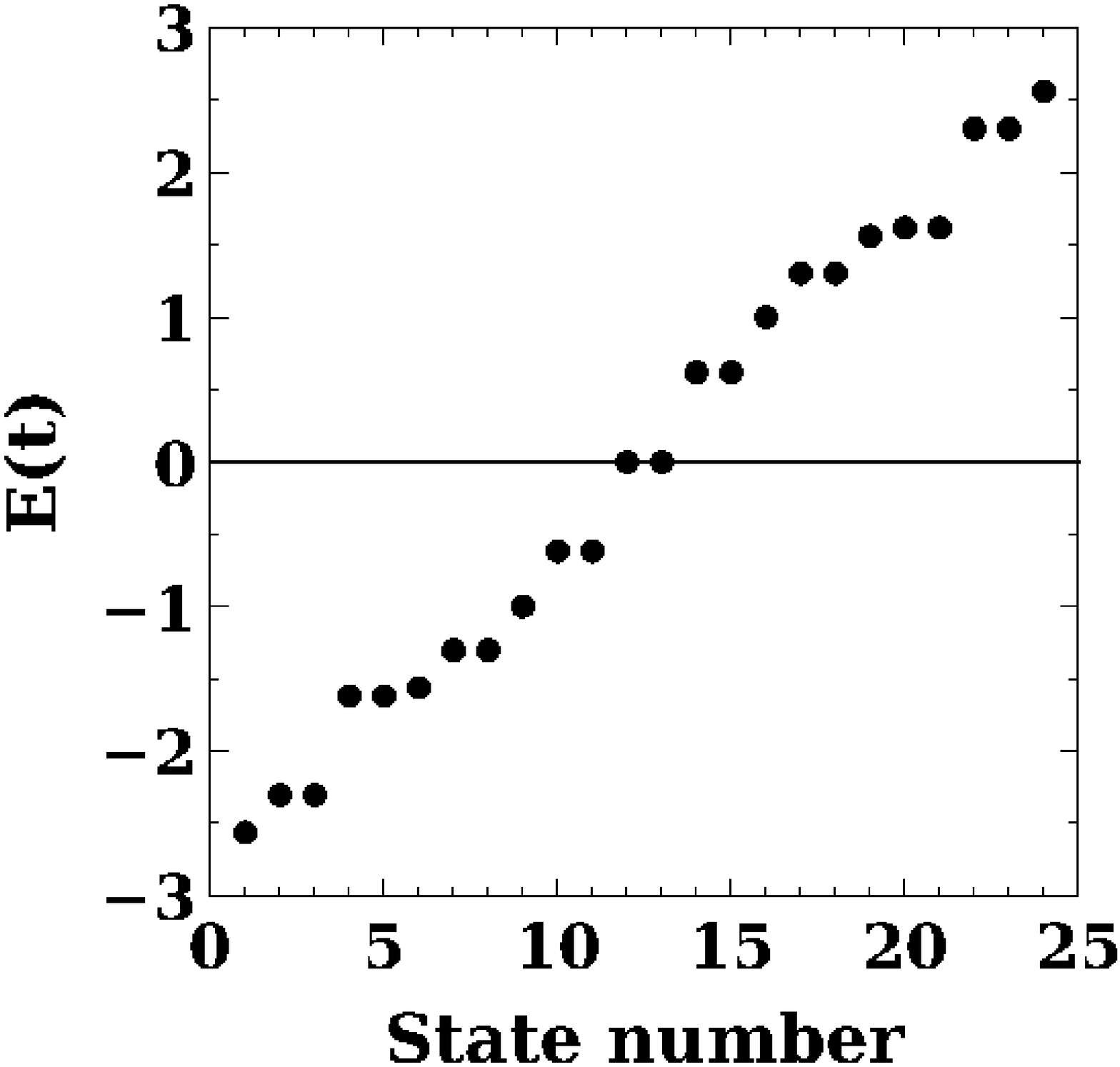}} & (b) \scalebox{0.25}{\includegraphics{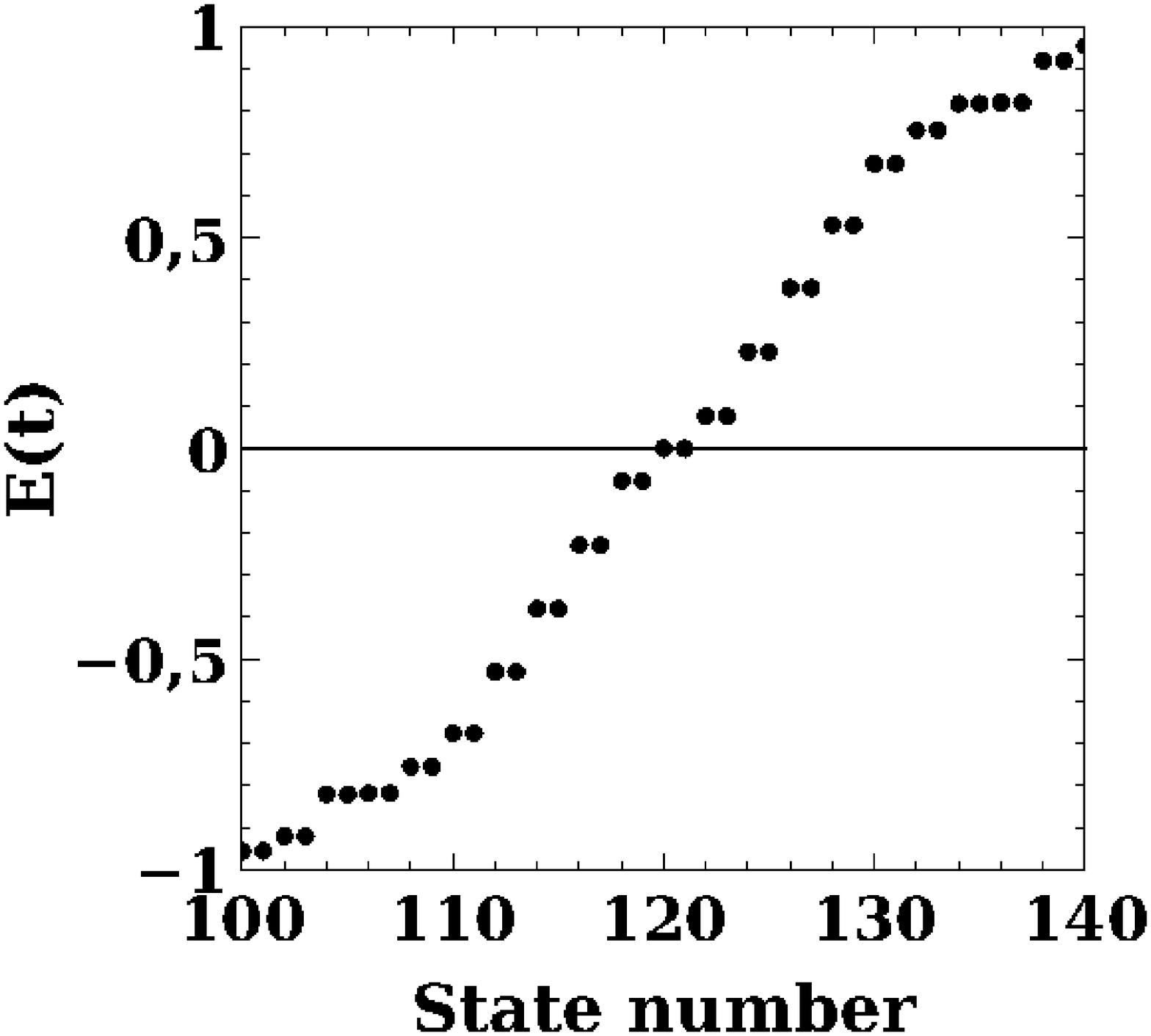}} & 
(c) \scalebox{0.1}{\includegraphics{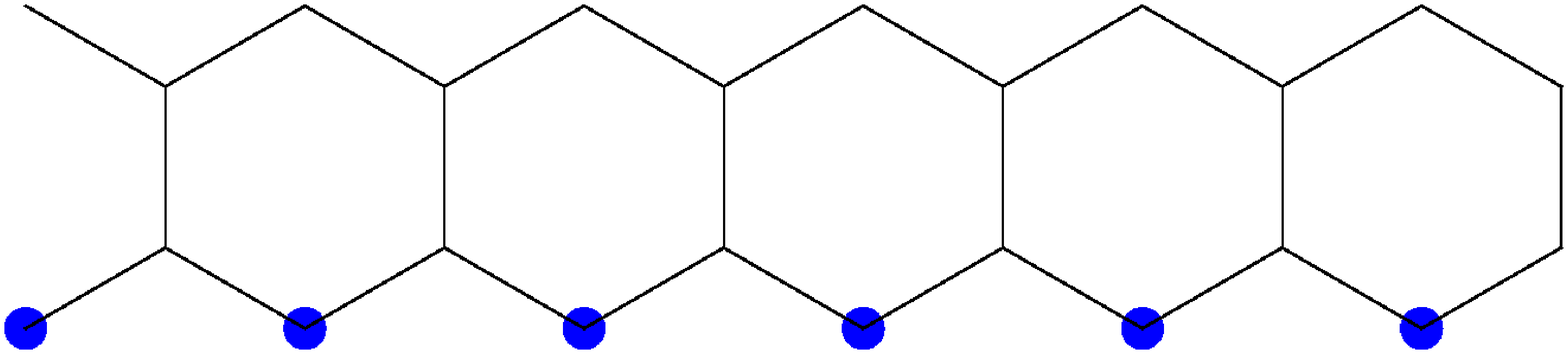}} & (d) \scalebox{0.25}{\includegraphics{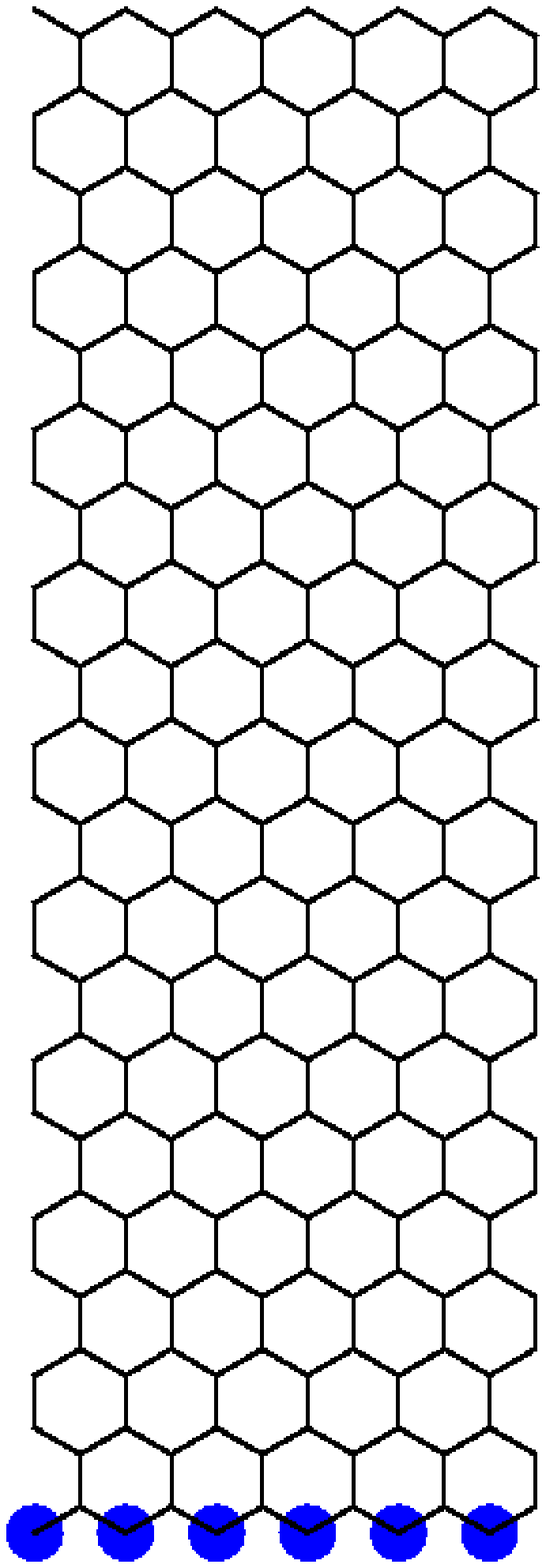}} \\
(e) \scalebox{0.25}{\includegraphics{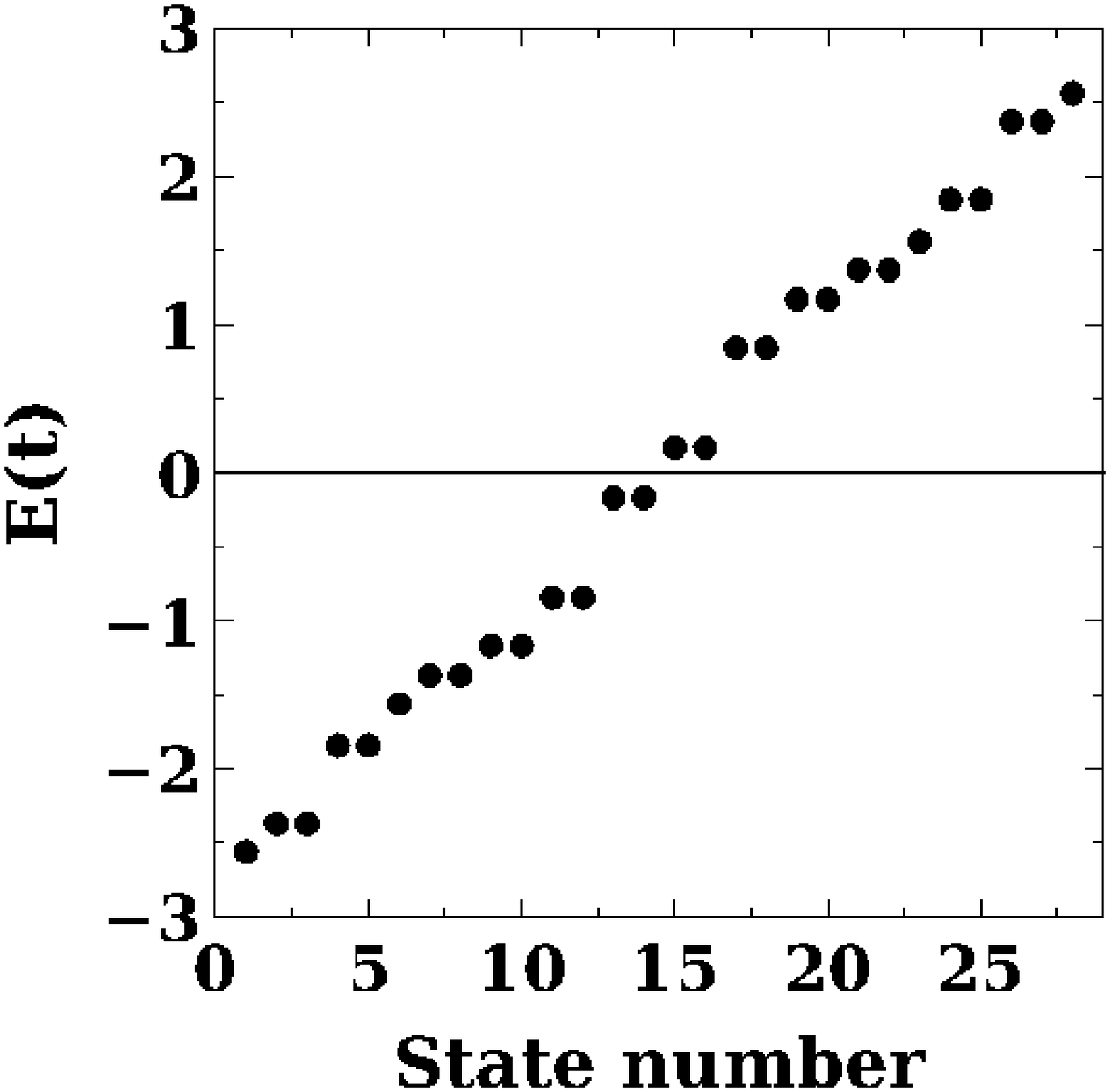}} & (f) \scalebox{0.25}{\includegraphics{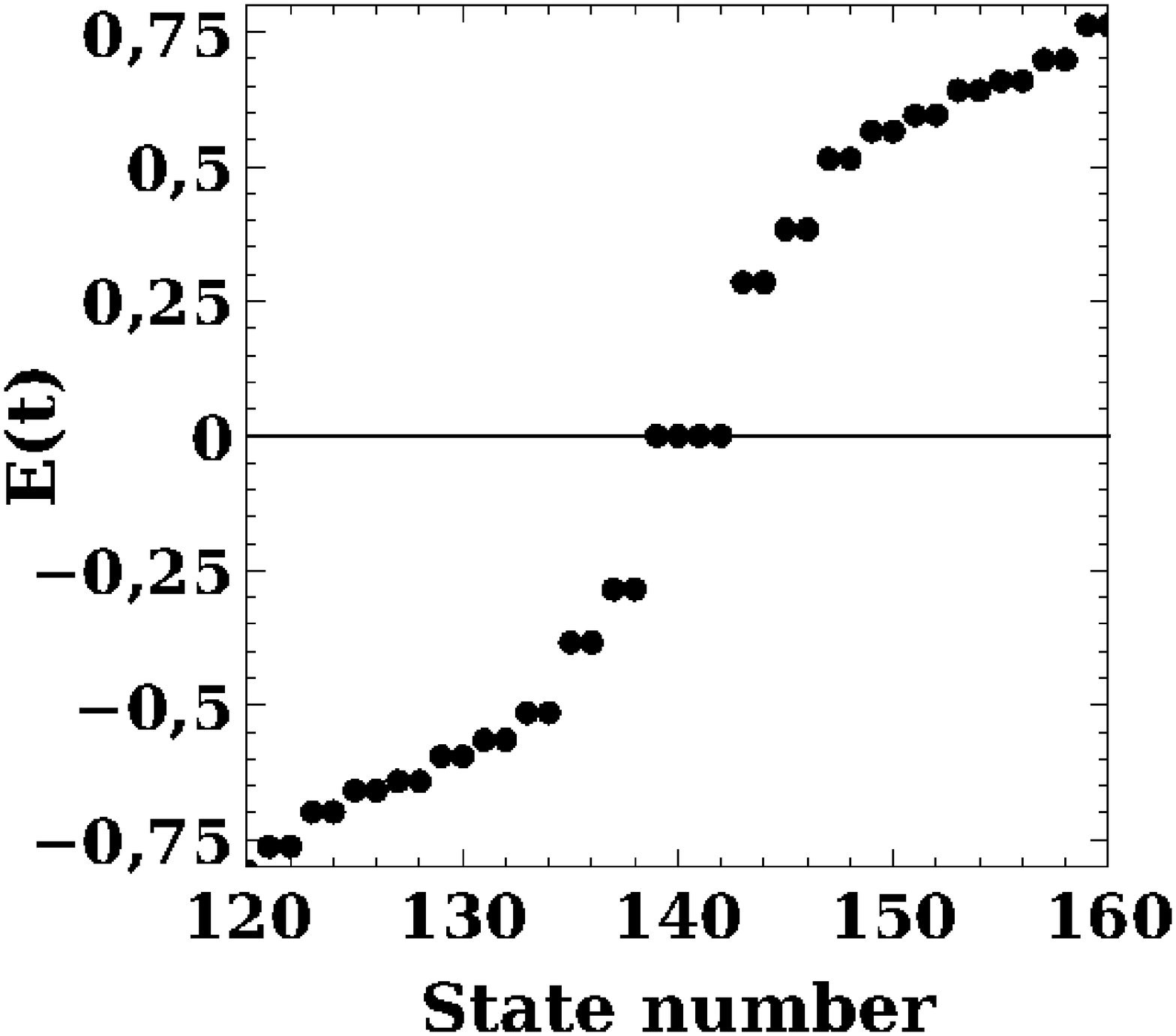}} & 
(g) \scalebox{0.1}{\includegraphics{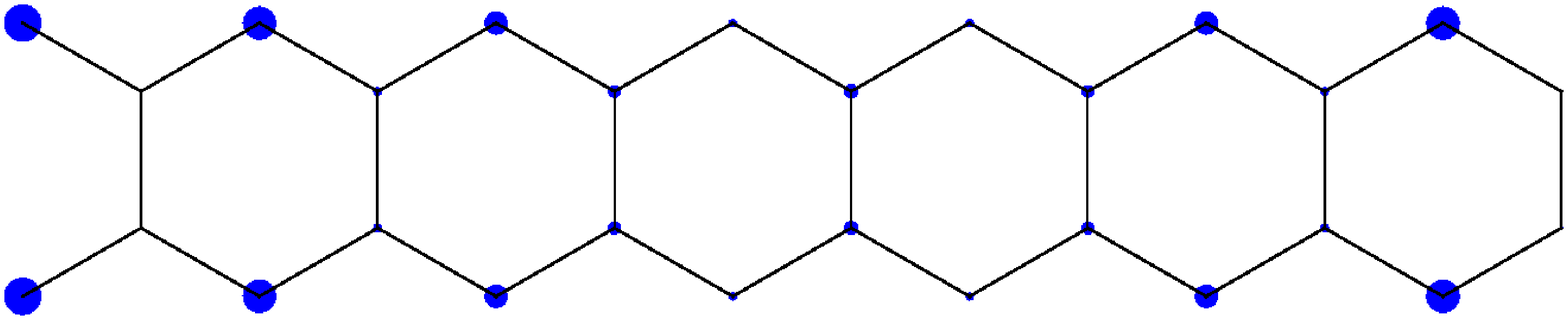}} & (h) \scalebox{0.25}{\includegraphics{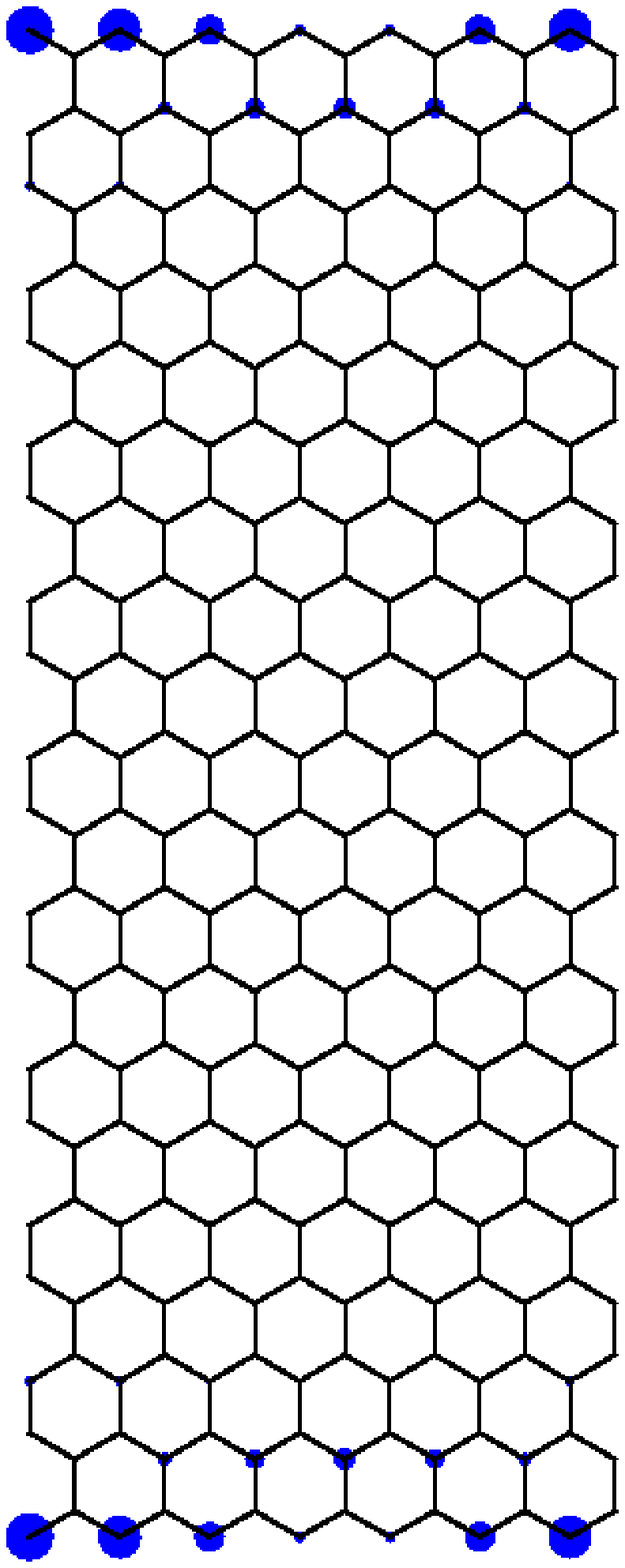}} \\
\end{tabular}
\caption{Top: Tight-binding eigenvalue spectra for [$6$]CC with (a) $L=1$, (b) $L=19$, as well as the plot of one of their corresponding zero-energy modes --(c) and (d)--. 
Bottom: Tight-binding eigenvalue spectra for [$7$]CC with (e) $L=1$, (f) $L=19$, as well as the plot of one of their corresponding ingap states --(g) and (h)--.
\label{fig:TB-spectra}}
\end{center}
\end{figure}
\end{landscape}

We first studied Cyclacenes under the Tight-Binding (TB) approximation. This permits to  connect with previous work about the electronic structures of short-sized 
CNTs\cite{bulusheva1998,correa2010}, and it provides a simple yet intuitive picture of what one would expect next employing more sophisticated theoretical methods. We 
consider a tight-binding model with one $\pi$ orbital per carbon atom, with first neighbour hopping $t$. The single-particle energy levels computed with this model for 
cyclic polyenes are shown in Figure~\ref{fig:TB-spectra} for  four structures with different values of $L$ and $n$. We remark the following results: (i) For even values 
of $n$, the eigenvalue spectra has two zero-energy modes, with each one strongly localized at one of the zigzag edges regardless of values of $L$; and (ii) when $n$ is 
odd and $L = 1$, there are no strict zero-energy modes, but two energy split doublets. We have verified that the the splitting decreases dramatically with $L$ (see 
Figure~\ref{fig:TB-spectra}). 
\\

The existence of both zero energy states and "quasi" zero energy states can be rationalized as follows. Structures with $L=1$ can be thought as 2 rings with $2n$ carbon atoms 
each. The inter-ring coupling occurs only between sites with the same parity, either odd or even, and fortunately the tight-binding problem of a single ring can be solved 
analytically. For a single ring with $N=2*n$ sites, the amplitude of the single-particle state on the site $\ell$ of the ring is $\psi_k (\ell)= e^{ik \ell}$. The eigenvalue 
associated with such $\psi_k$ is $\epsilon(k)= 2 t \cos k=2t \cos (\frac{2\pi m}{N})$. The permitted values of $k$ are obtained by imposing that the amplitude satisfies 
$\psi_k (\ell)=\psi_k (\ell+N)$ from which we infer $k=\frac{2m\pi}{N}$ where $0,\pm1, \pm2 ...$ The number of states is thus given by $N$. If we now look for zero energy 
states, $\epsilon(k)=0$, this happens when $(\frac{2\pi m}{N})=\frac{\pm \pi}{2}$, or $m=N/4$. Therefore, for even $n$ values, $N=2n$ is divisible by 4 and there are two zero 
modes in the ring. If $n$ is odd, $2N$ is not divisible by 4, and there are not zero modes in the ring.
\\

We now discuss how to build the zero modes of the $L=1$ structures, based on the zero modes of the even $n$ rings. For that matter, we realize that the wave functions 
$\psi_z(\ell)= \frac{1}{\sqrt{N}}e^{\pm i\frac{\pi}{2} \ell}$ can now be combined leading to symmetric and antisymmetric states, giving two orthogonal $E=0$ states living in 
the odd and even sites of the ring, respectively. When the two rings are coupled to form the $L=1$ structure with even $n$, the zero modes hosted in the sites that are 
{\em not} affected by the inter-ring coupling  are not perturbed, keeping $E=0$. In contrast, the other pair of zero modes, hosted in sites that are affected by inter-ring 
interactions, form molecular states with energy $\pm t$ (see Figure~\ref{fig:TB-spectra}a).
\\

The previous argument can be extended to larger systems, e.g. $L=3$ formed by fusing two $L=1$ blocks. Before their coupling, each $L=1$ block has 2 zero modes. The coupling 
affects the carbon atoms where the zero energy modes are hosted, and two zero modes hosted are consequently hybridized, but two zero modes hosted at the top and bottom edge of 
the structure survive. Therefore, regardless the value of $L$, the structure has two sublattice polarized $E=0$ zero modes (see Figure S1 in the Supporting Information). This 
is a non-trivial result, since the existence of 
$E=0$ modes is secured when there are more atoms in one of the two triangular sublattices, which is not the case of these systems. The ultimate origin of these zero modes relies 
on the existence of a structural symmetry that commutes with the sublattice operator\cite{koshino2014}. Therefore, local distortions would lift the degeneracy of the zero mode 
doublet.
\\

We now consider the case of odd-$n$ structures, for which the elementary rings do not have zero energy modes. Instead, they have 2 doublets that overshoot/undershoot the 
$k=\pm\frac{\pi}{2}$ condition, with energy close to, but strictly different from zero. When fused into the $L=1$ structure, these states give also 4 low energy modes. However, 
as $L$ increases, keeping fixed $n$, the energy of these 4 states gets closer to $E=0$. We can understand this starting from the limit of a infinite graphene ribbon with 
zigzag edges and width $L$. It is well known\cite{nakada1996} that the resulting energy bands have two quasi-flat bands with energy close to $E=0$, that vanish exactly at 
$k= \pm {\pi}{a}$, where $a$ is the unit cell length. The energy levels of our finite size nanotubes can be considered as a sampling in the $k$ space of the spectrum of these 
ribbons. For even-$n$ structures, the sampling is such that it contains the $k=\pm \frac{\pi}{a}$ doublet. For odd-$n$ structures, the sampling is such that it misses that point, 
leading to a quartet formed by the two energy bands and the two $k$ points closest to  the zone boundary. Note that the latter four states can be linearly combined to obtain 
four sublattice polarized zero-energy modes; two at one zigzag edge, and two at the other edge (see for instance our recent work\cite{ortiz2018}). Increasing $n$ and $L$ will 
result in a larger number of quasi 
zero energy modes. The case of [$n$]CPH was also analyzed, see Figure S2 in the Supporting Information, without displaying any zero-energy modes in this case. 
These results from the tight-binding model qualitatively agree with the diradical and tetraradical character of even and odd ciclacenes, respectively, disclosed previously 
and confirmed quantitatively by previous multiconfigurational wavefunction calculations by some of the authors\cite{perez-guardiola2018}. 
\\

\subsection{Energy magnitudes for increasingly longer nanorings}

\begin{figure}
\begin{center}
\scalebox{0.5}{\includegraphics{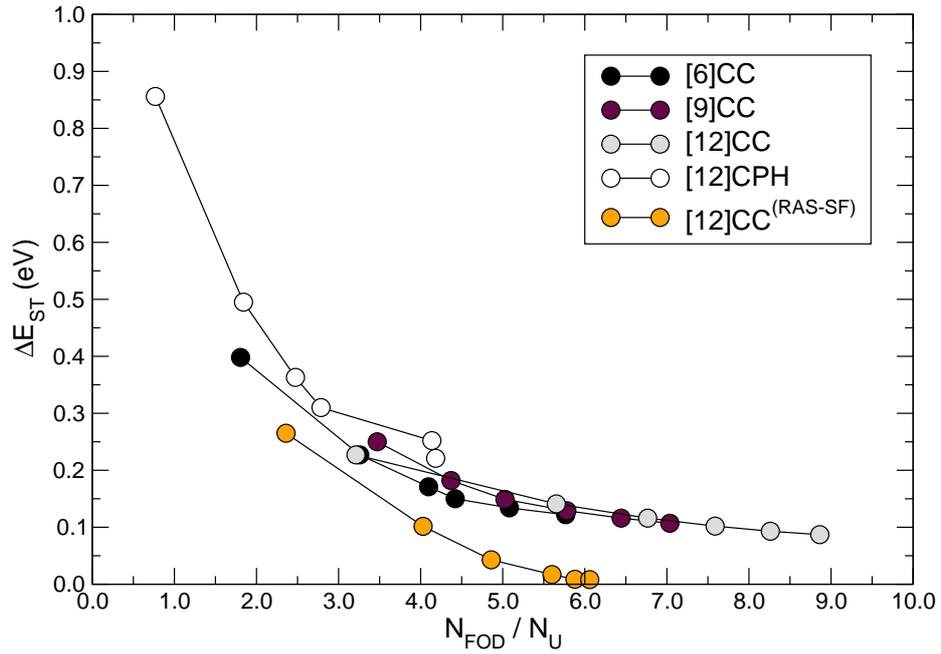}} \\
\caption{Evolution of the $\Delta E_{\mathrm{ST}}$ values obtained from the FT-DFT method, as a function of the N$_{\mathrm{FOD}}$ values for for all systems ($L=1-6$). 
The \mbox{N$_{\mathrm{U}}$} results for [$12$]CC ($L=1-6$) systems are also shown. The straight lines are a guide to the eye.
\label{fig:DE_ST_vs_N_FOD}}
\end{center}
\end{figure}

\begin{figure}
\begin{center}
\scalebox{0.5}{\includegraphics{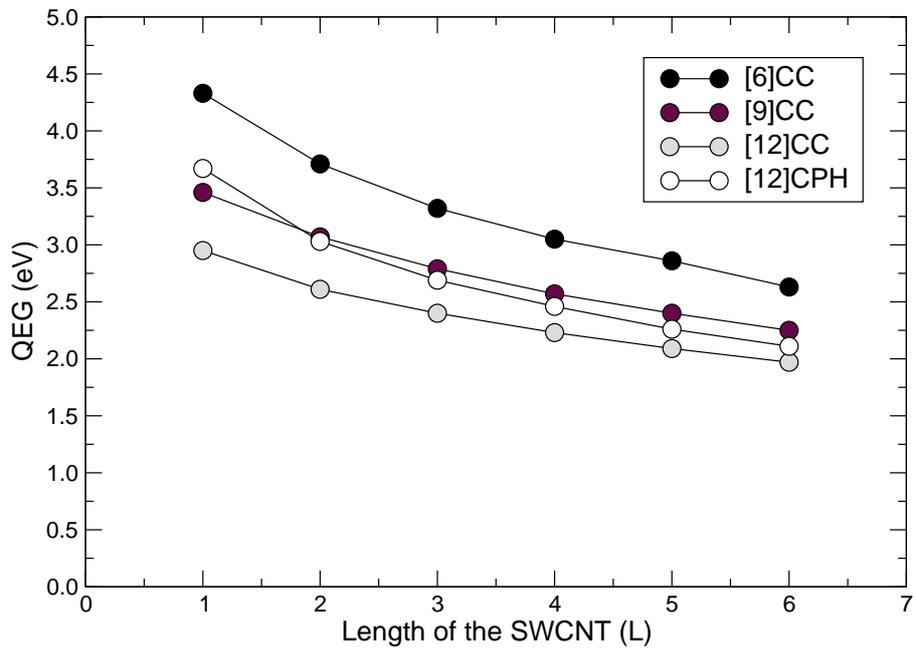}} \\
\caption{Evolution of the QEG values obtained from the FT-DFT method, as a function of the nanotube size ($L=1-6$) for all systems.
\label{fig:QEG_vs_N_p}}
\end{center}
\end{figure}

As key energy magnitudes for understanding the underlying electronic structure of these systems, we choose the Singlet-Triplet energy difference 
($\Delta E_{\mathrm{ST}}$), the Vertical Ionization Potential (VIP), the Vertical Electron Affinity (VEA), and the Quasi-Energy Gap (QEG), defined respectively as:
\begin{eqnarray}
\label{eqn:5}
\Delta E_{\mathrm{ST}} & = & ^3E_N -~^1E_N, \\
\mbox{VIP}             & = & ^2E_{N-1} -~^1E_N, \\
\mbox{VEA}             & = & ^1E_N -~^2E_{N+1}, \\
\mbox{QEG}             & = & \left( ^2E_{N+1} -~^2E_{N-1} \right) -~^1E_N,
\end{eqnarray}
with $^SE_N$ being the FT-DFT calculated total energy of the $N$-electron system with spin multiplicity $S$, and $N+1$/$N-1$ indicating the corresponding charged systems. The 
VIP and VEA energies, or simply $I$ and $A$ in the following for simplicity, relate with the chemical potential ($\mu$) and other indicators to charge donation and 
acceptance\cite{gazquez2007}:
\begin{eqnarray}
\label{eqn:6}
\mu      & = & -\frac{I + A}{2}, \\
\omega   & = & \frac{\left( I + A \right)^2}{4\left( I - A \right)}, \\
\omega^- & = & \frac{\left( 3I + A \right)^2}{16\left( I - A \right)}, \\
\omega^+ & = & \frac{\left( I + 3A \right)^2}{16\left( I - A \right)},
\end{eqnarray}
with $\omega$ the electrophilicity, and $\omega^-$ ($\omega^+$) the electrodonating (electroaccepting) power. Table 1 gathers the results for all the systems considered, with 
the main findings summarized as follows:

\begin{itemize}

\item The N$_{\mathrm{FOD}}$ values increase with the nanotube length ($L$) and diameter ($n$), reaching significantly high values for the longest oligomer of [$n$]CC, and tend 
to saturate with values of $L$, with slight differences between odd and even nanotubes. These values indicate their polyradical nature and the growing multiconfigurational 
character of their singlet ground state. Comparing [$12]$CC and [$12$]CPH of the same length, we can observe how the values for the latter are roughly halved with respect to the 
former systems, which agrees with the higher stability of the latter arising from the larger number of aromatic rings according to Clar's rule\cite{esser2015}.

\item The $\Delta E_{\mathrm{ST}}$ values decrease with the nanotube length ($L$) and diameter ($n$), reaching values as low as $0\mathrm{.}1-0\mathrm{.}2$ eV unless for 
the case of [$12$]CPH oligomers. The results of Figure~\ref{fig:DE_ST_vs_N_FOD} reveal a clear correlation between the singlet-triplet energy difference and the N$_{\mathrm{FOD}}$ 
values, a trend previously disclosed for linear and cyclic oligoacenes\cite{perez-guardiola2018}. Note how this relationships also qualitatively prevails when 
RAS-SF/6-31G(d) calculated $\Delta E_{\mathrm{ST}}$ and N$_{\mathrm{FOD}}$ values are instead considered, for the [$12$]CC oligomers taken as example 
(see Figure S2).

\item The VIP (VEA) values slightly decrease (increase) with the nanotube length ($L$), roughly comprising between $5-6$ ($2-3$) eV and monotonically converging with the 
system size, as it was also previously observed for cyclic oligoacenes\cite{wu2016}. The behaviour of the QEG values, or fundamental gap, is reminiscent of those trends found for 
VIP and VEA, with their evolution shown in Figure~\ref{fig:QEG_vs_N_p}. Note that values are expected to converge towards a limit found between $2\mathrm{.}0-2\mathrm{.}5$ eV, which 
is considerably lower than that found for other hoop-shaped conjugated molecules\cite{liu2015}.

\item The reactivity indexes considered are related to the maximum electron flow when a molecule is embedded into a bath at a constant potential $v(\mathbf{r})$, and complement 
the information provided by VIP and VEA for those cases where approximately more or less than one electron can be transferred\cite{geerlings2003}. Actually, the values for all 
systems increase with the nanotube length ($L$), and follow the sequence $\omega^- > \omega > \omega^+$ independently of $L$ and $n$, thus indicating a propensity to donate 
rather than to accept charge. The ratio $\frac{\omega^-}{\omega^+} \simeq 2-3$, with the highest (lowest) ratio found for [$6$]CC ([$12$]CC) with [$9$]CC and [$12$]CPH keeping 
intermediate values. 
%Interestingly, mono- and direduced states of [$6$]CPP were recently synthesized and crystallized\cite{spisak2018} opening the use of these systems for multielectron storage.

\end{itemize}

This set of results also relates with the energetic stability of finite size tubes, which should ideally correlate with experimental abundance of SWCNT. Fixing the value 
of $L=6$ for [$6$]CC, [$9$]CC, [$12$]CC, and [$12$]CPH, which also allows a qualitative comparison with previous theoretical estimates for all the combinations $n+n = 8-18$ 
for both $(n,0)$ and $(n,n)$ cases\cite{hedman2015}, one can see how the FOD-based descriptors (i.e. N$_{\mathrm{FOD}}$) and energy magnitudes (i.e. $\Delta E_{\mathrm{ST}}$) 
would predict the following stability: [$12$]CPH $>$ [$6$]CC $>$ [$9$]CC $>$ [$12$]CC, in agreement with the preference for armchair and near-armchair abundance of fragment of 
SWCNT in the initial steps of their growth\cite{hedman2017}.
\\

\subsection{Topology of the FOD density and occupation numbers for increasingly longer nanorings}

\begin{landscape}
\begin{figure}
\begin{center}
\begin{tabular}{cccc}
\scalebox{0.14}{\includegraphics{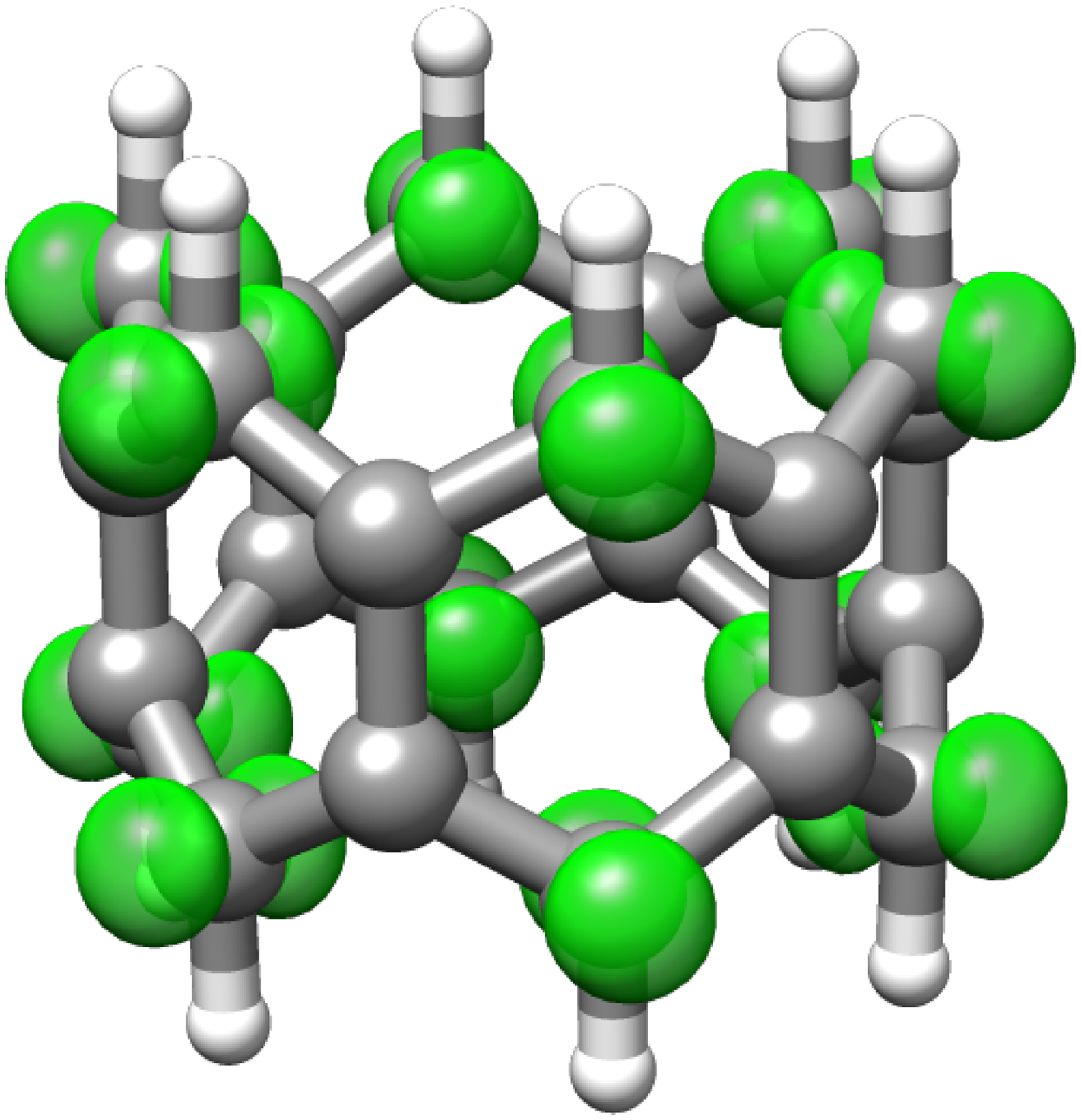}} & \scalebox{0.14}{\includegraphics{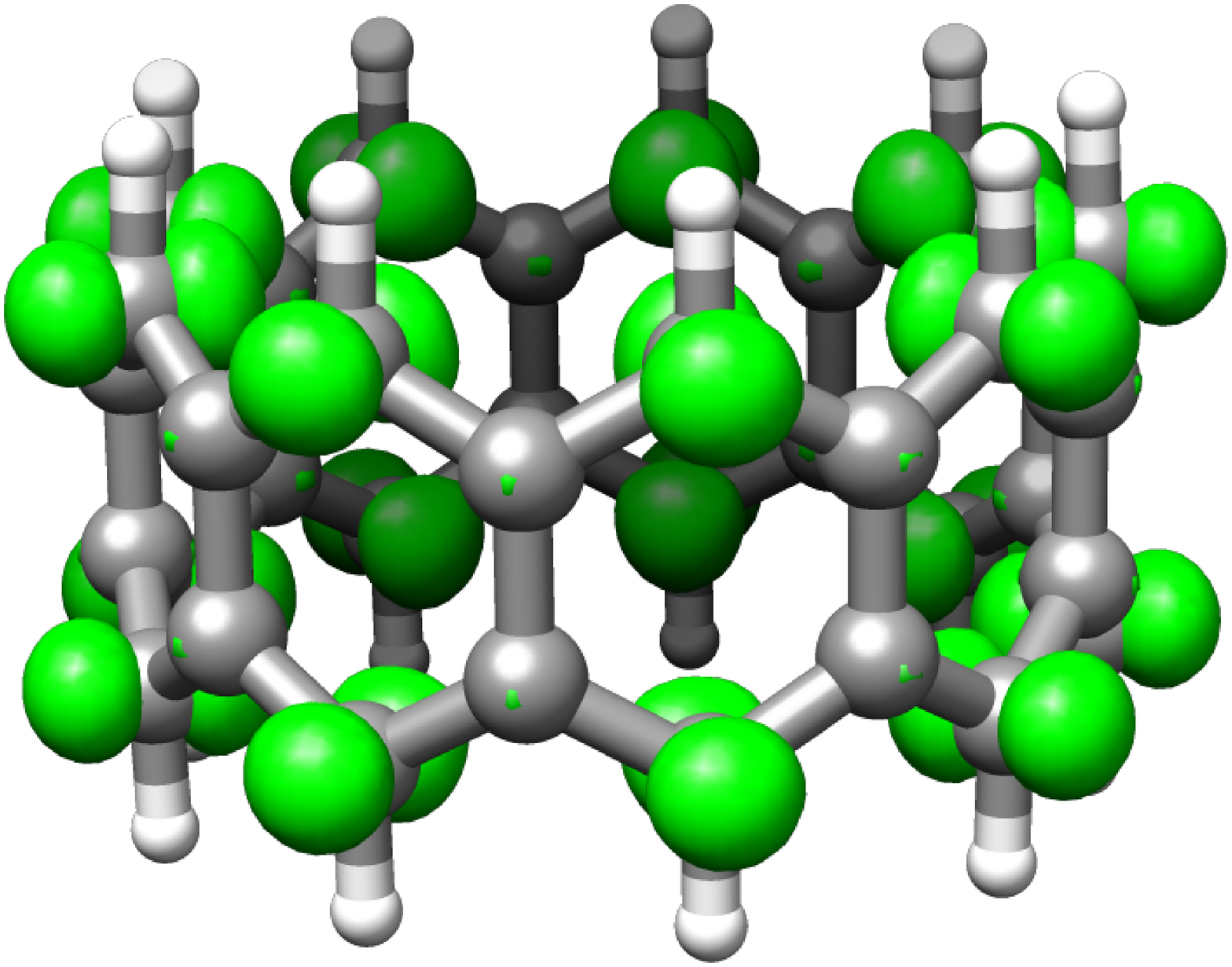}} & \scalebox{0.2}{\includegraphics{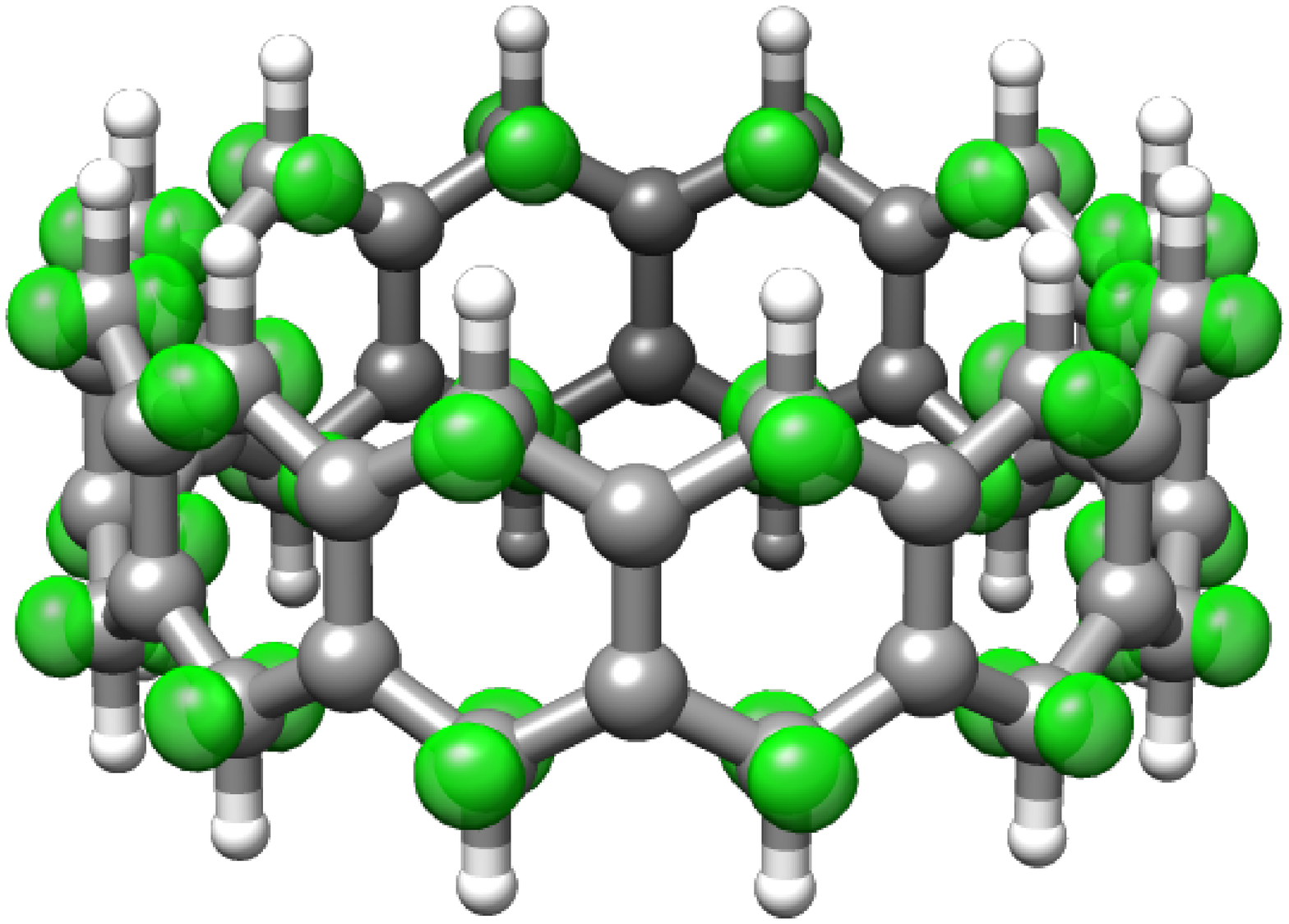}} & \scalebox{0.2}{\includegraphics{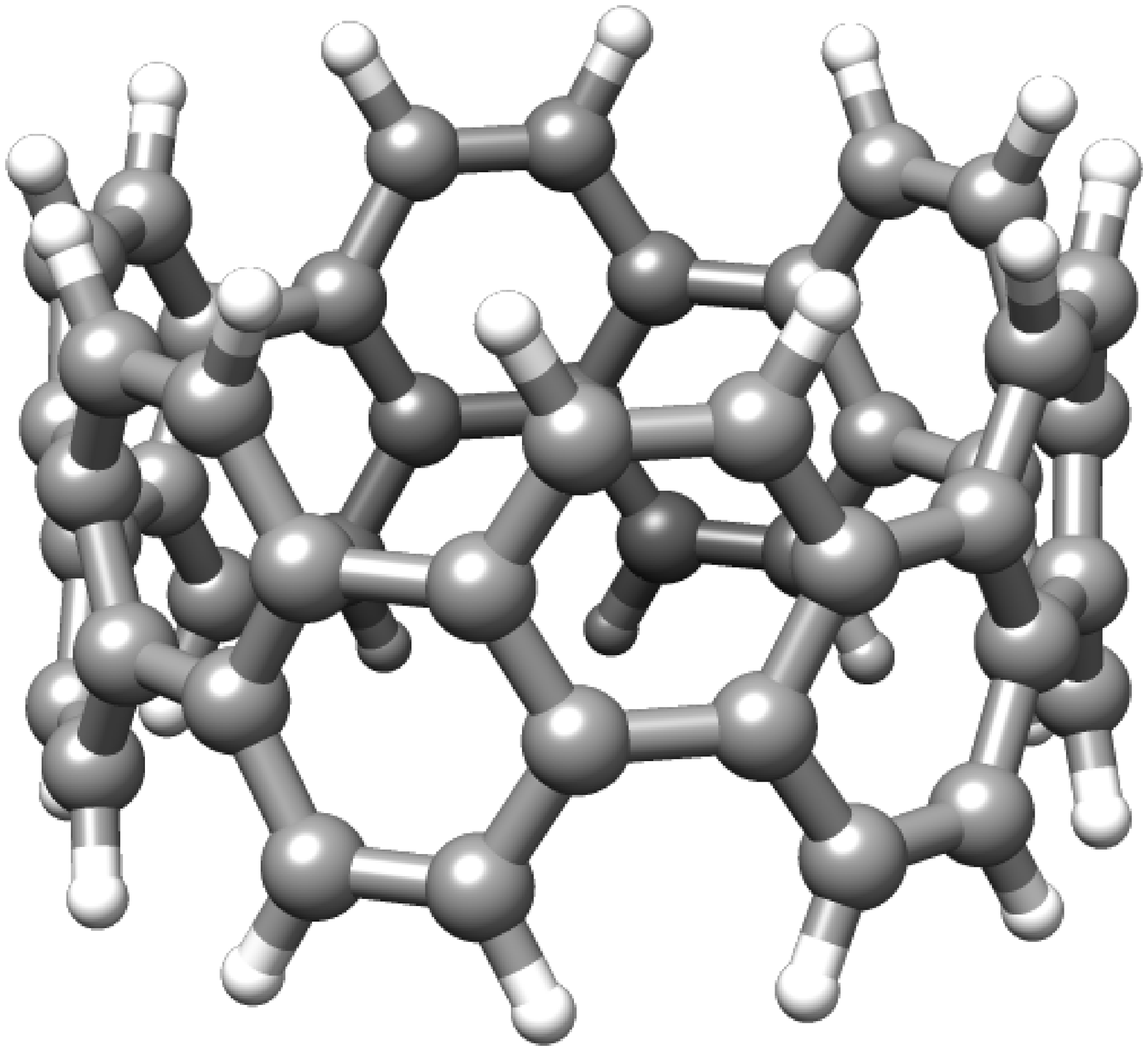}} \\
\scalebox{0.29}{\includegraphics{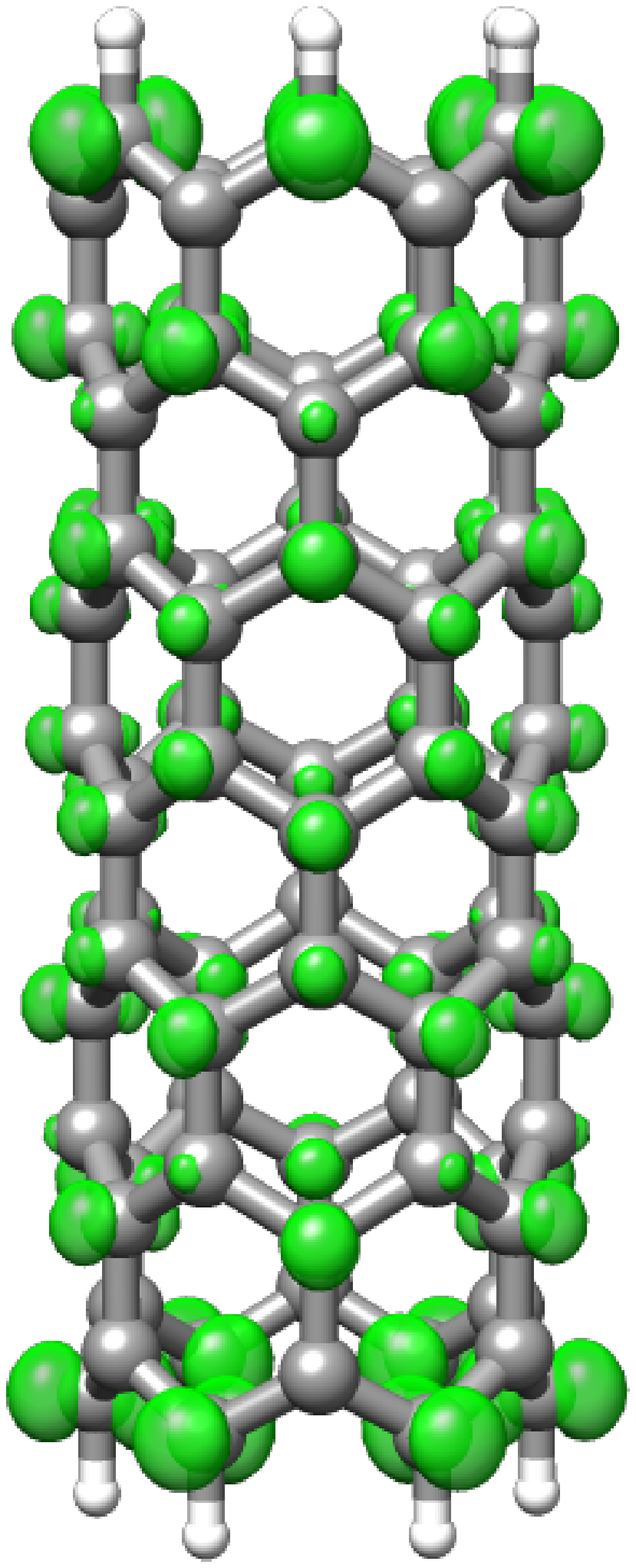}} & \scalebox{0.31}{\includegraphics{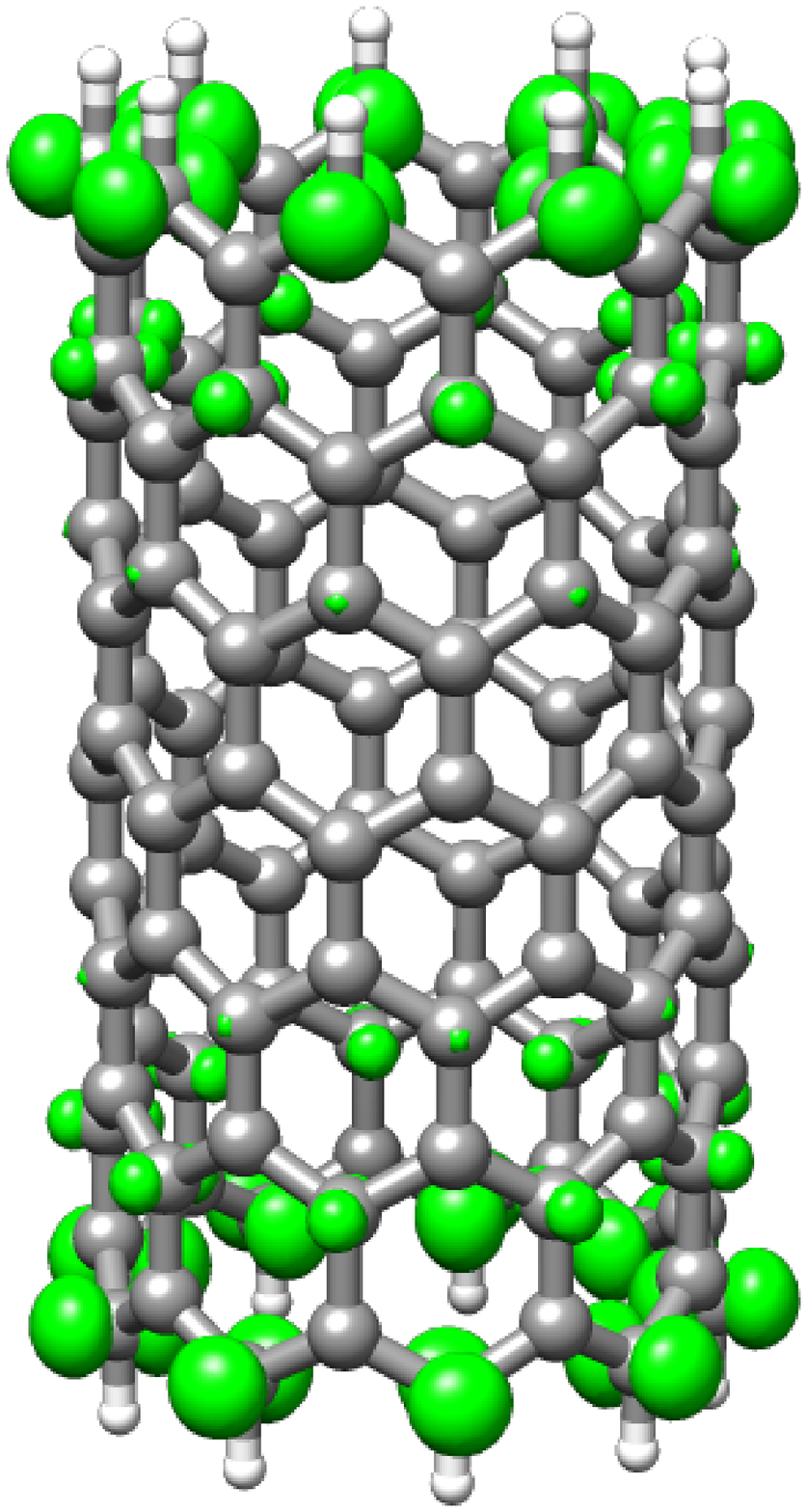}} & \scalebox{0.3}{\includegraphics{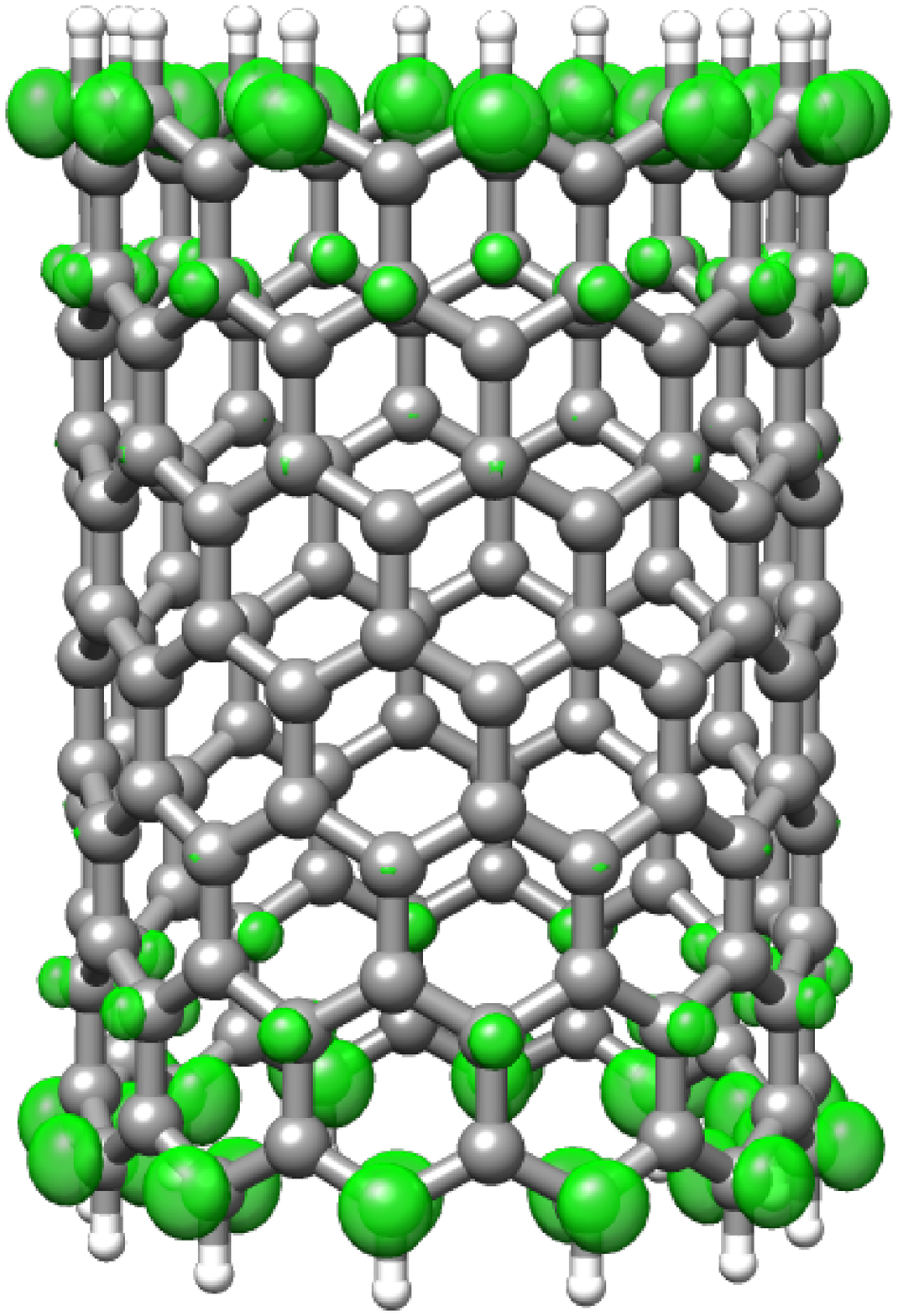}} & \scalebox{0.32}{\includegraphics{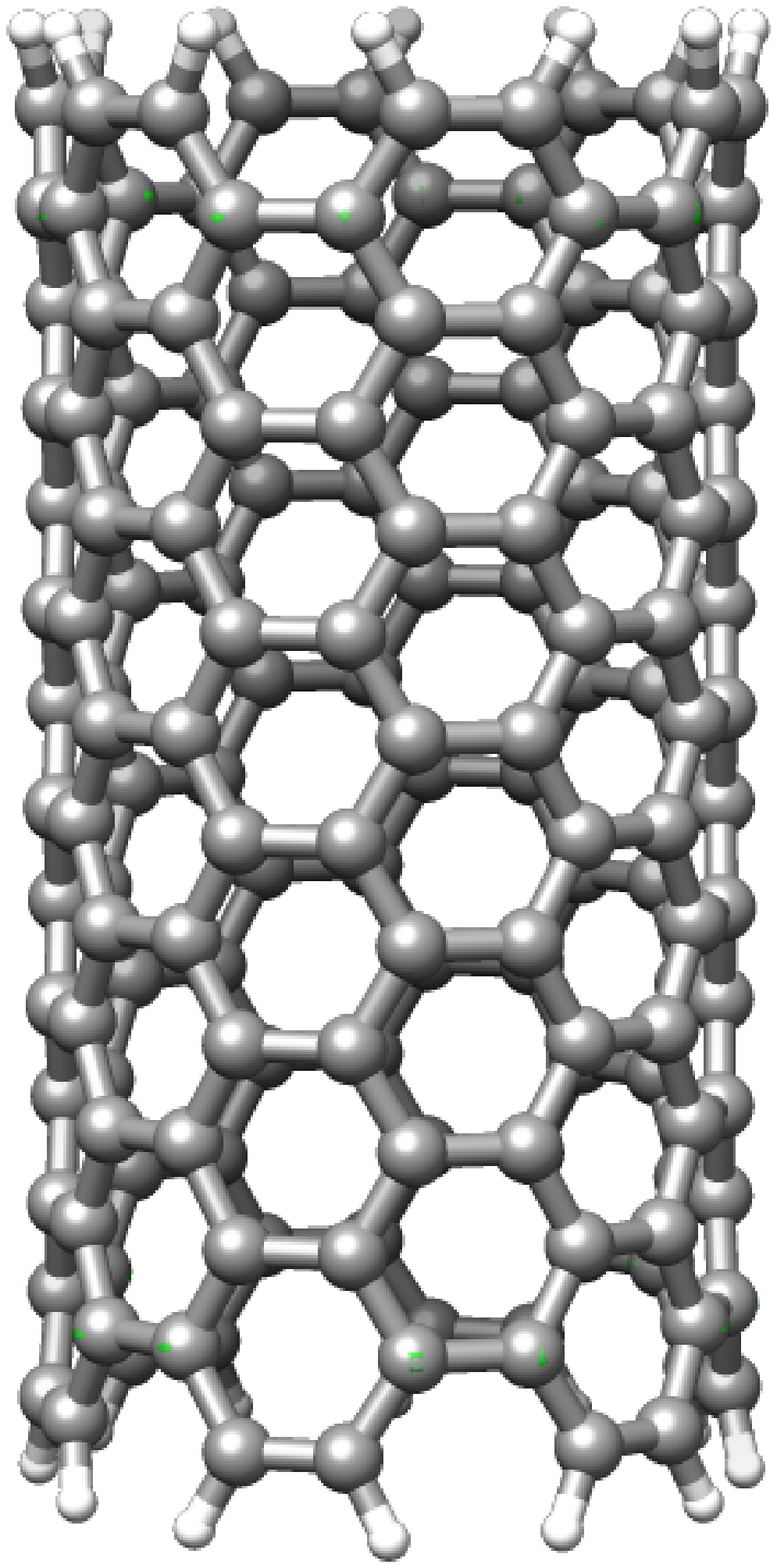}} \\
\end{tabular}
\caption{Chemical structures and plots ($\sigma = 0\mathrm{.}005$ e/bohr$^3$) of the FOD density obtained from the FT-DFT method, for the shortest (top) and longest (bottom) 
oligomer of the set of [$n$]CC ($n=6,9,12$) and [$12$]CPH compounds (from left to right).}
\label{fig:density_FOD}
\end{center}
\end{figure}
\end{landscape}

\begin{figure}
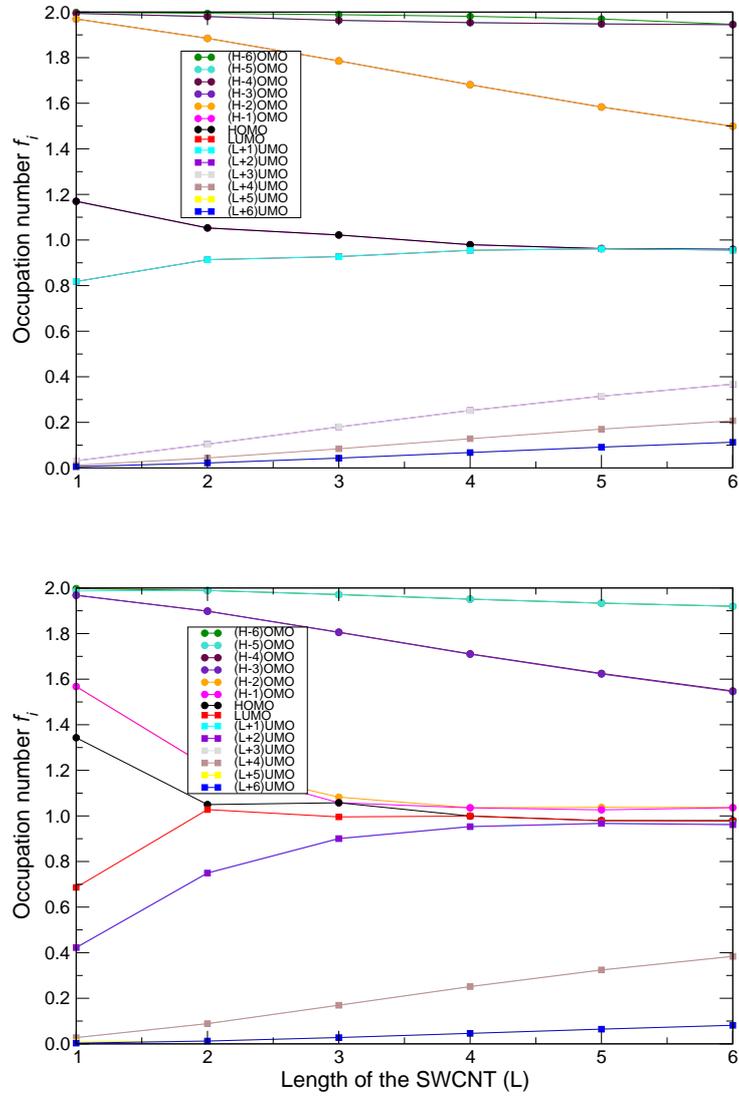

\begin{center}
\scalebox{0.4}{\includegraphics{Nocc_9CC.eps}} \\ \vspace{.8cm}
\scalebox{0.4}{\includegraphics{Nocc_12CC.eps}} \\
\caption{Evolution of $f_i$ values for [$9$]CC (top) and [$12$]CC (bottom), as obtained from the FT-DFT method, for the (H-6)OMO to (L+6)UMO set of orbitals.
\label{fig:f_i}}
\end{center}
\end{figure}

We represent in Figure~\ref{fig:density_FOD} the spatial distribution of $\rho^{\mathrm{FOD}}({\mathbf r})$, as obtained from FT-DFT calculations, for the shortest ($L=1$) and 
longest ($L=6$) oligomer of all systems considered so far: [$6$]CC, [$9$]CC, [$12$]CC, and [$12$]CPH. Note that when $L \rightarrow \infty$, [$n$]CC and [$n$]CPH systems will lead 
to zigzag $(n,0)$ and armchair $(n,n)$ SWCNTs, with the oligomers considered here transitioning between nanobelts and nanorings, respectively, and the infinitely extended systems 
thoroughly studied in the literature. Note also that for computational studies of finite size SWCNT we need to saturate the edges with H atoms. For the [$6$]CC oligomers, it is 
immediately seen how the density mostly locates at the edges and primarily on the non-bridging C atoms. Going from $L=1$ to $L=6$ (see Figures S3-S6 in the Supporting Information) 
extends the distribution of $\rho^{\mathrm{FOD}}({\mathbf r})$ to every ring and to all C atoms within them, while keeping the adherence of (or propensity of finding) the density 
at the edges. When the diameter of the oligomers increases, i.e. going from [$6$]CC to [$9$]CC or to [$12$]CC, the density concentrates almost exclusively on the edges, essentially 
keeping the same pattern as before regarding on which C atoms is located. However, the set of [$12$]CPH oligomers behave completely different, and actually one would need a lower 
cut-off ($\sigma = 0\mathrm{.}002$ e/bohr$^3$) to make the $\rho^{\mathrm{FOD}}({\mathbf r})$ visible, see Figure S7 in the Supporting Information, indicating a weaker 
polyradicaloid character in agreement with the larger $\Delta E_{\mathrm{ST}}$ values obtained with the FT-DFT method. Interestingly, the spatial distribution of the 
RAS-SF $\rho^{\mathrm{FOD}}({\mathbf r})$ density, see Figure S8 for the set of [$12$]CC oligomers, closely resembles the ones obtained before.
To further illustrate the increase of N$_{\mathrm{FOD}}$ with $L$ for [$n$]CC systems, we plot the fractional occupation numbers of a sufficiently large window of orbitals
in Figure~\ref{fig:f_i} for the representative [$9$]CC and [$12$]CC cases. The Highest Occupied Molecular Orbital (HOMO) and Lowest Unoccupied Molecular Orbital (LUMO)
represent the frontier orbitals, the number of fractionally occupied orbitals distribute around them, showing an increase of the (poly-)radical character with the nanotube size.
\\

\begin{figure}
\begin{center}
\scalebox{0.4}{\includegraphics{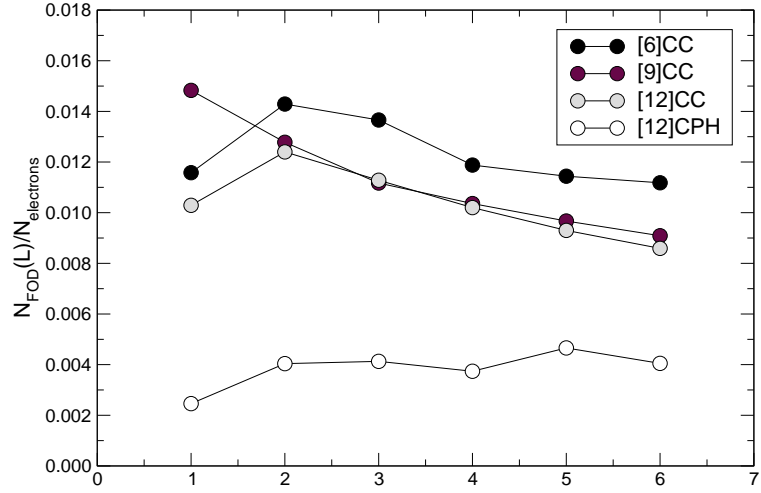}} \\ \vspace{.8cm}
\scalebox{0.4}{\includegraphics{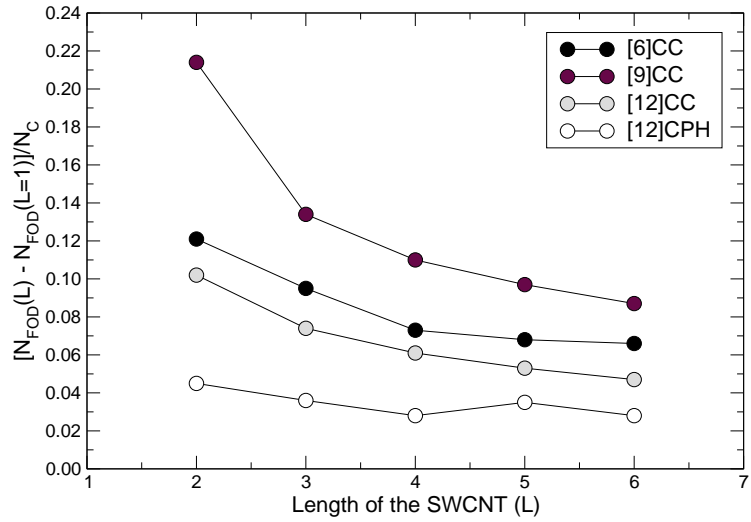}} \\
\caption{Evolution of N$_{\mathrm{FOD}}$ values as obtained from the FT-DFT method, renormalized by the number of electrons (top) and by attenuating edge effects (bottom) 
for each nanotube size ($L=1-6$).
\label{fig:N_FOD}}
\end{center}
\end{figure}

To investigate if the increase of N$_{\mathrm{FOD}}$ values with the nanotube length is due to cumulative or intrinsic effects, we represent in Figure~\ref{fig:N_FOD} 
the total values divided by the number of electrons of each system: $\frac{\mbox{N$_{\mathrm{FOD}}(L)$}}{\mbox{N}_{\mathrm{electrons}}}$. We can easily see how for 
both [$n$]CC and [$n$]CPH systems each C atom contributes approximately the same to the polyradicaloid nature of the systems, although more pronouncedly in the case of [$n$]CCs, 
with a slight trend to decrease (increase) for [$n$]CCs ([$n$]CPHs) as a function of the nanotube length $L$. In a second step, we renormalize the N$_{\mathrm{FOD}}$ values by 
subtracting the values for the shortest oligomer ($L=1$) to the rest of the values, and dividing by the number of C atoms (N$_{\mathrm{C}}$), as given by:
\begin{equation}
\label{eqn:7}
\mbox{normalized~N$_{\mathrm{FOD}}$} = \frac{\mbox{N$_{\mathrm{FOD}}$}(L) - \mbox{N$_{\mathrm{FOD}}$}(L=1)}{\mbox{N$_{\mathrm{C}}$}},
\end{equation}
highlight the importance of edge effects, i.e. how the inner rings contribute to the global N$_{\mathrm{FOD}}$ values, which we also represent in Figure~\ref{fig:N_FOD}. 
Interestingly, the edge effects seem to saturate with the system size for [$n$]CC systems, confirming that very long nanotube length would not suffer from infinitely 
higher and higher N$_{\mathrm{FOD}}$ values. Note also the nearly constant value of the normalized FOD values for the case of [$n$]CPH systems, in agreement with the 
homogeneous distribution of the FOD density along their structure (see Figure S7 in the Supporting Information). This finding also agrees with previous investigations 
as the length of the ($n,0$) and $(n,n)$ cases increases, even causing a switch to occur in the relative stability of zigzag vs. armchair structures\cite{hedman2017}.
\\

\subsection{The effect of end-capping of the longest nanorings}

We also investigate the tailored capping of systems such as [$6$]CC and [$12$]CC, both with a $L=6$ length as a matter of illustration, to disclose if the nature of the caps 
modify significantly or not 
the electronic properties of finite size nanotubes. We also note that the shape of a nanotube cap strongly depend on the base end, that is, whether a pentagon or a hexagon is 
situated at the cusp which actually determines the nature of the whole cap\cite{lair2006,melle2015}. Note that finite size end-capped nanotubes have been proposed as nanovehicles 
to deliver specific targets into cells after crossing their membranes\cite{hofinger2011}, and are reported to magnify the field-emission microscopy images\cite{tanaka2004}, 
opening new nanotechnological applications. 
\\

\begin{figure}
\begin{center}
\begin{tabular}{cc}
\scalebox{0.3}{\includegraphics{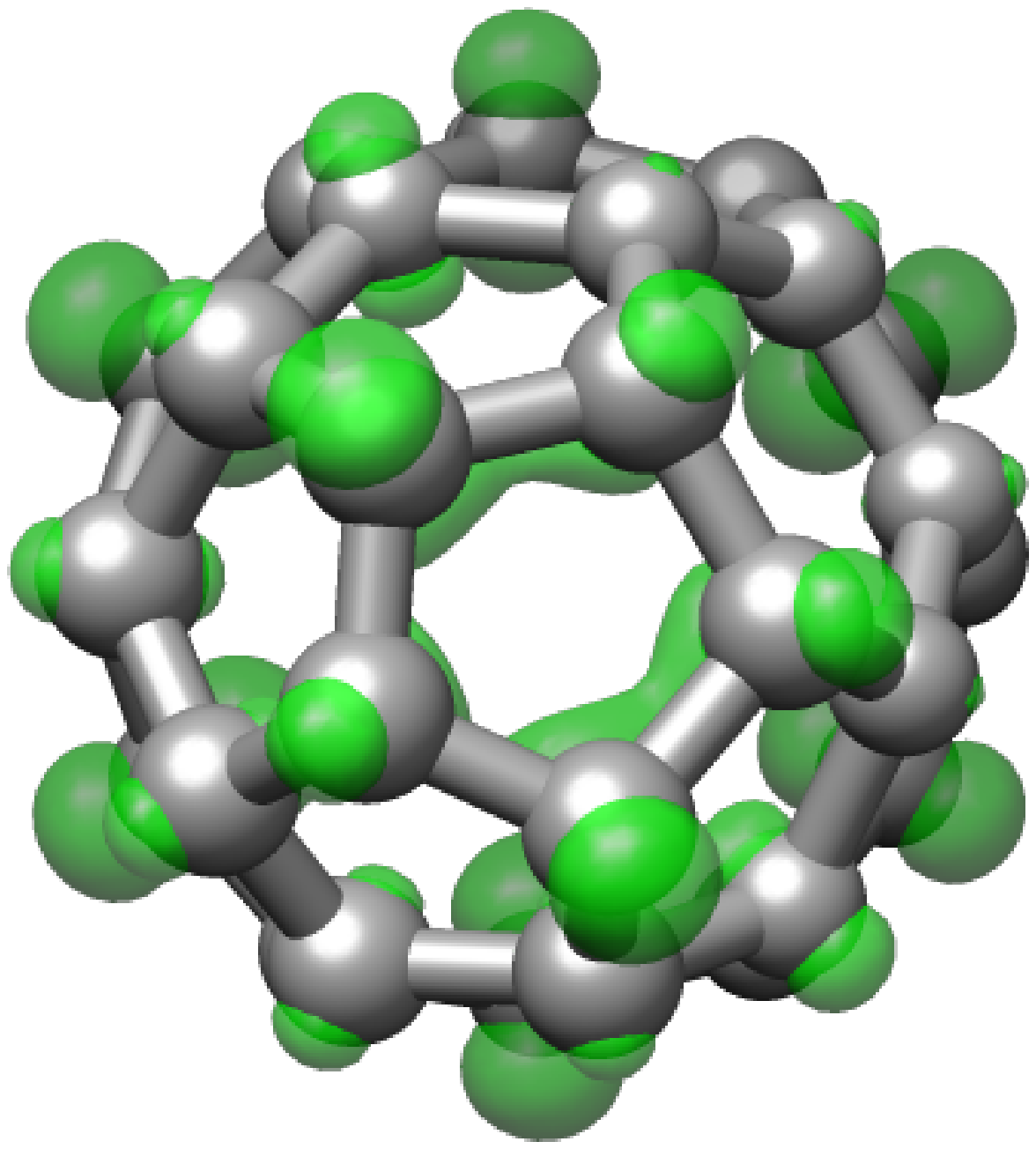}} & \scalebox{0.23}{\includegraphics{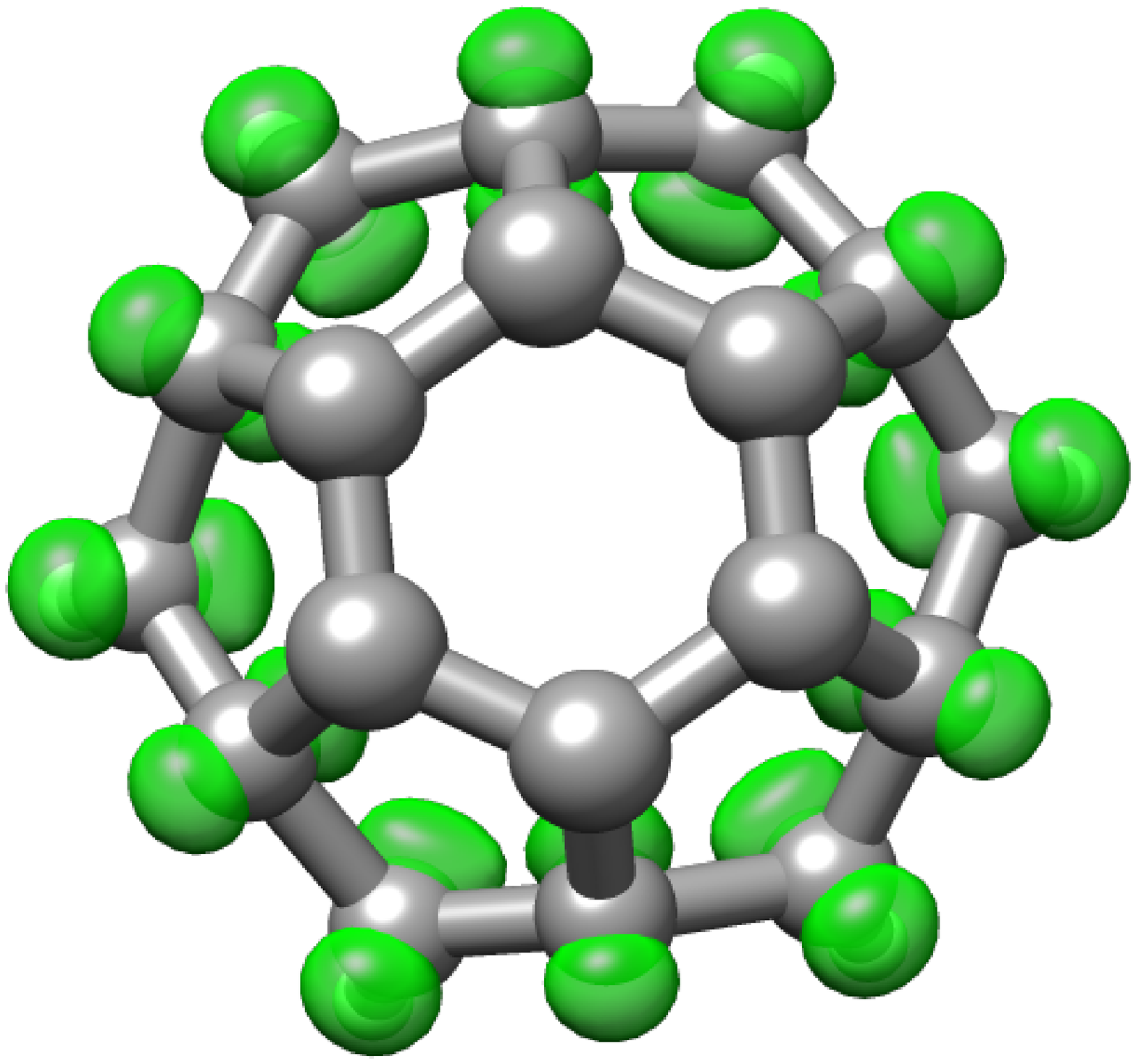}} \\
N$_{\mathrm{FOD}}$ = 1.76 & N$_{\mathrm{FOD}}$ = 1.36 \\ \\
\scalebox{0.3}{\includegraphics{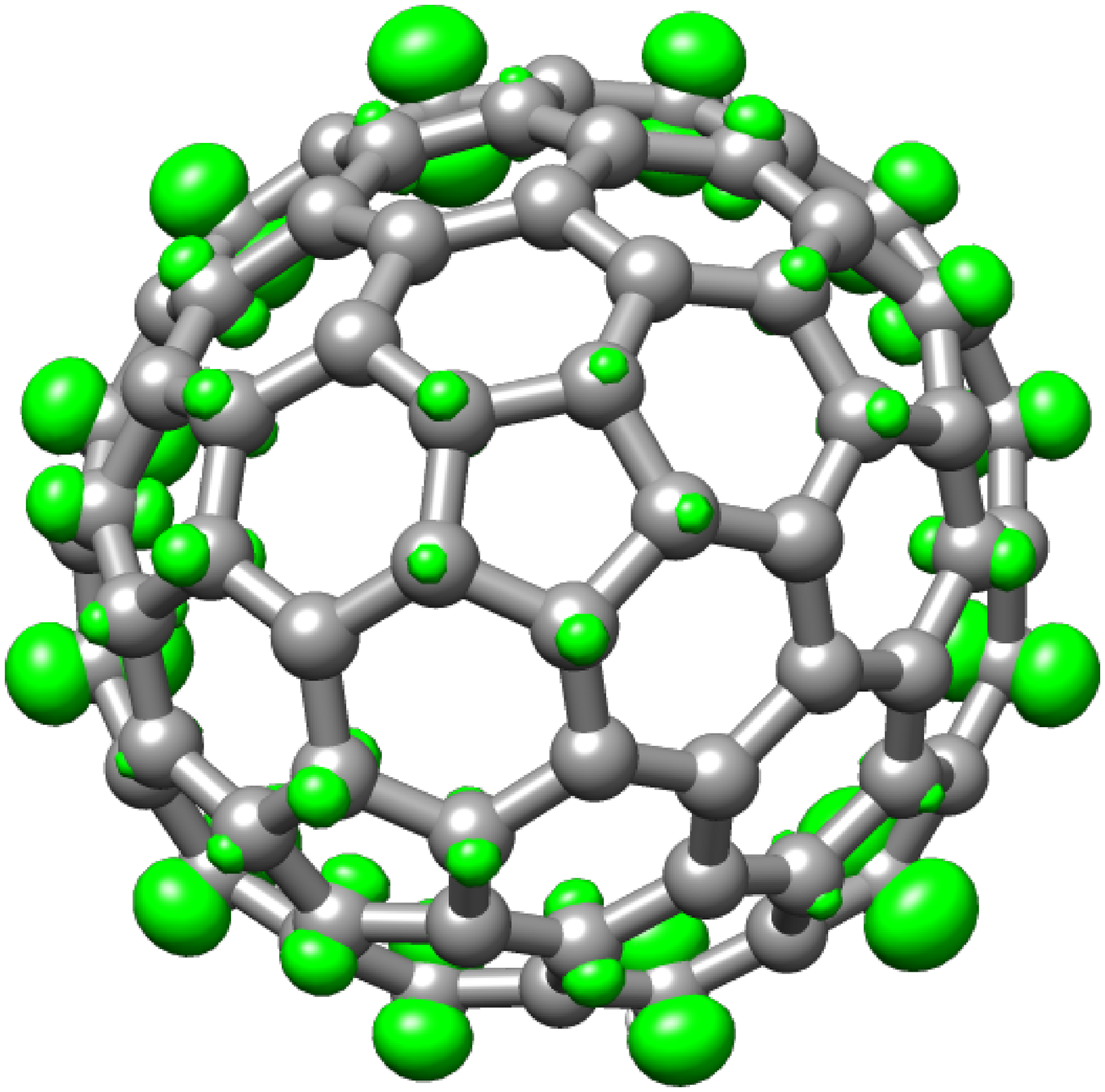}} & \scalebox{0.24}{\includegraphics{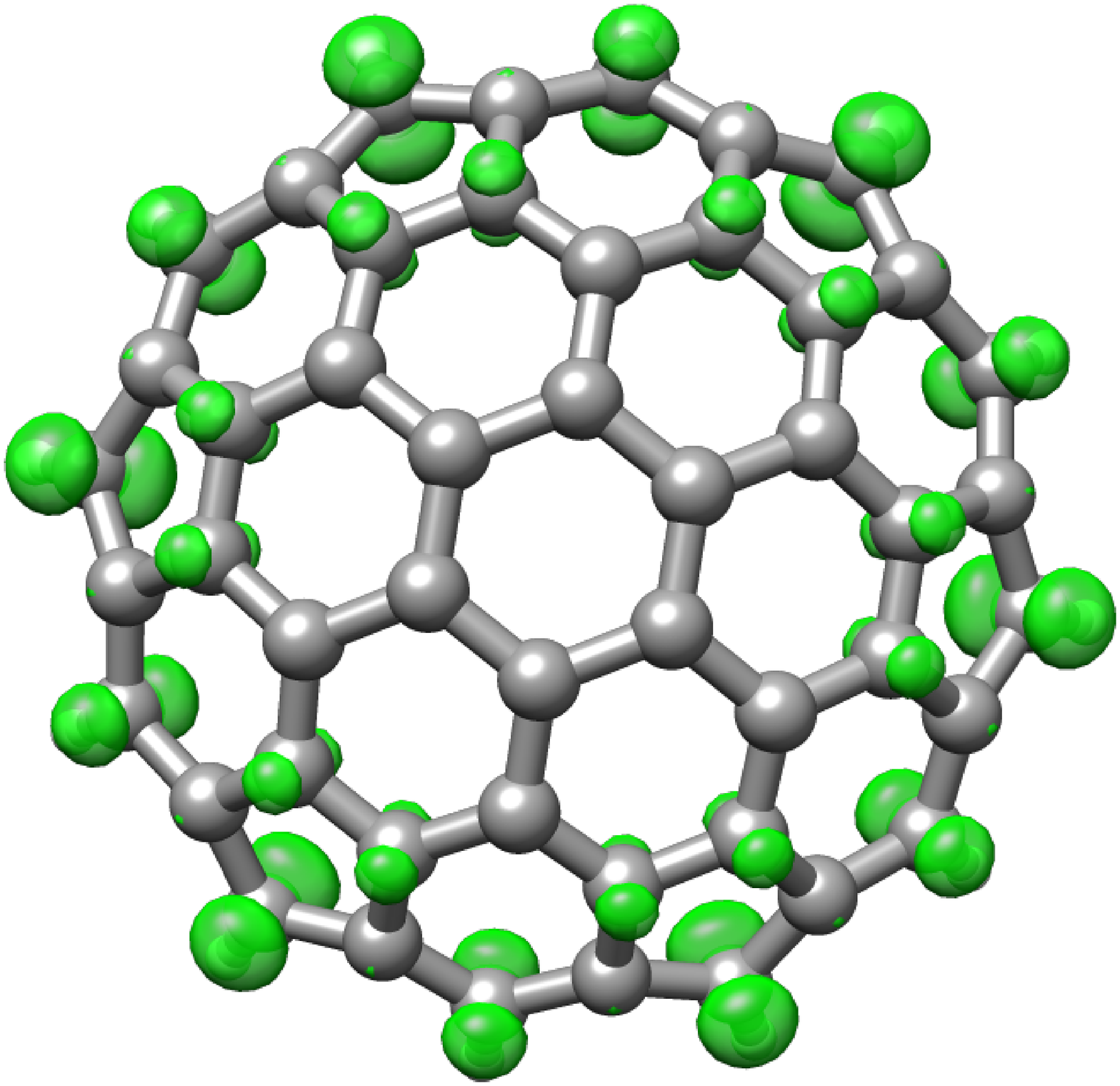}} \\
N$_{\mathrm{FOD}}$ = 3.92 & N$_{\mathrm{FOD}}$ = 3.18 \\
\end{tabular}
\caption{Chemical structures and plots ($\sigma = 0\mathrm{.}005$ e/bohr$^3$) of the FOD density obtained from the FT-DFT method for the caps with pentagonal (left) and 
hexagonal (right) base, for both [$6$]CC (top) and [$12$]CC (bottom) systems.
\label{fig:caps}}
\end{center}
\end{figure}

\begin{figure}
\begin{center}
\begin{tabular}{cc}
\scalebox{0.37}{\includegraphics{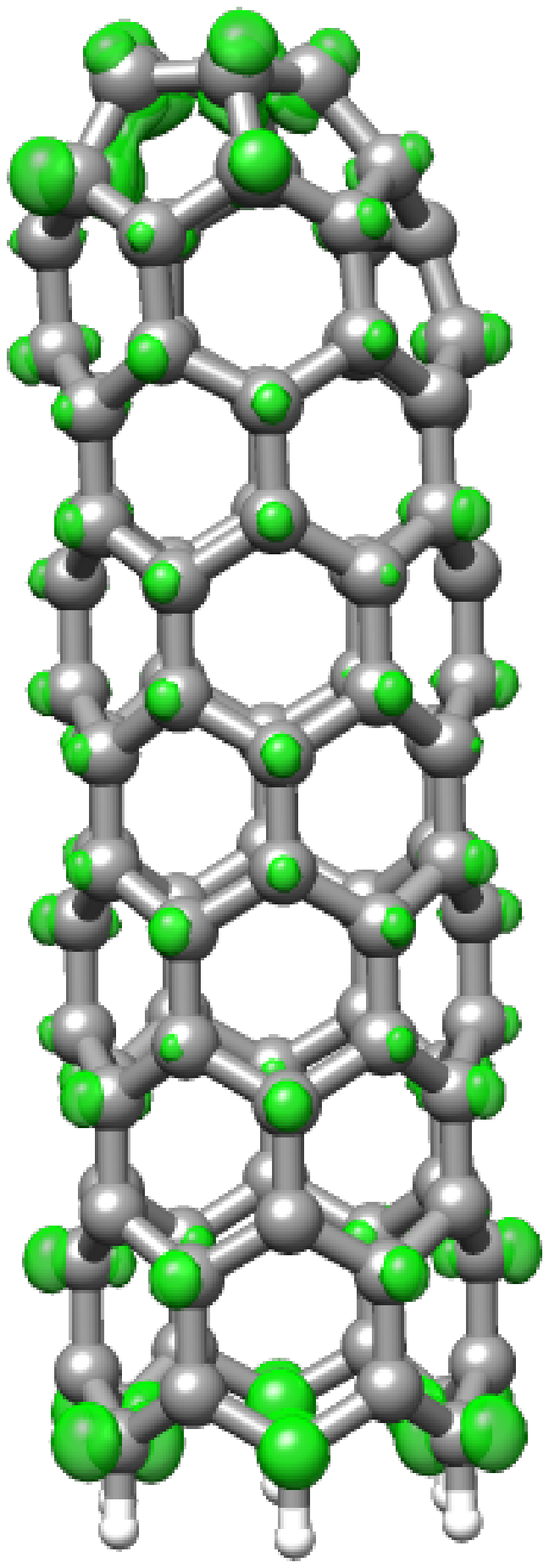}} & \scalebox{0.3}{\includegraphics{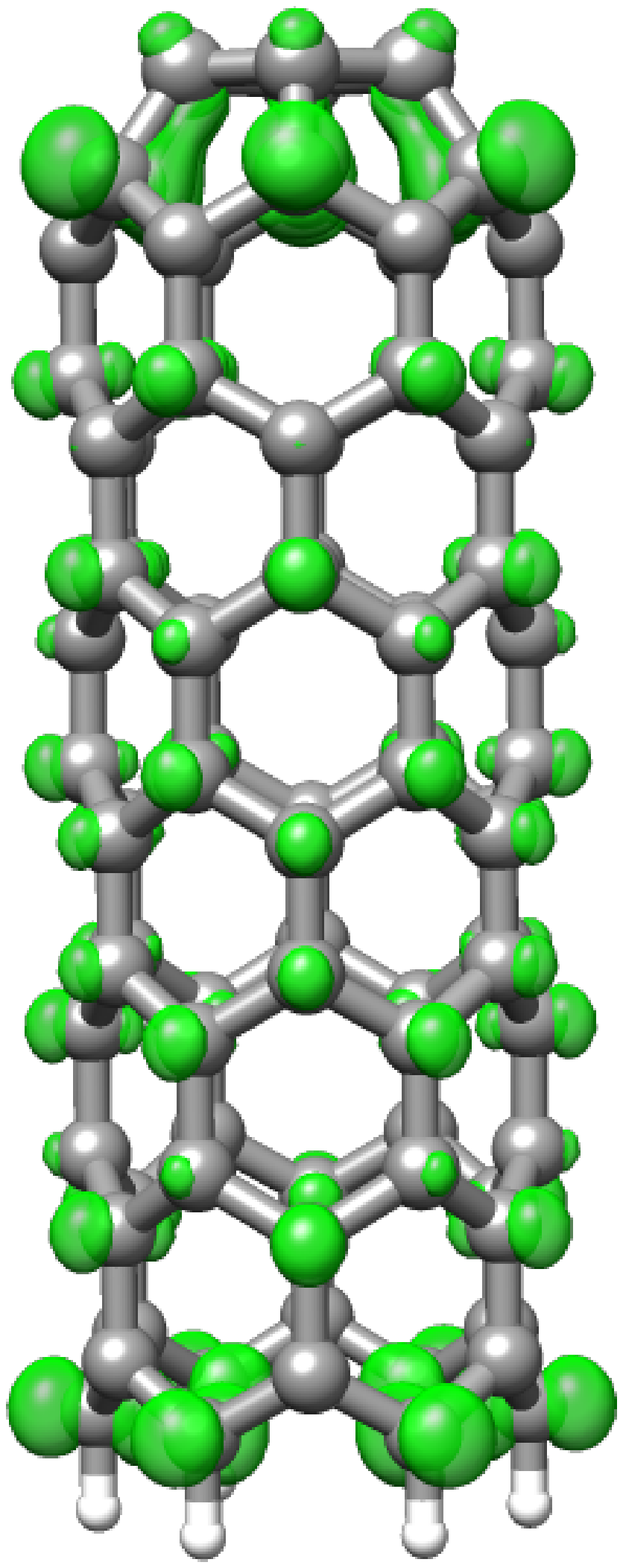}} \\
N$_{\mathrm{FOD}}$ = 6.37 & N$_{\mathrm{FOD}}$ = 5.98 \\ \\
\scalebox{0.34}{\includegraphics{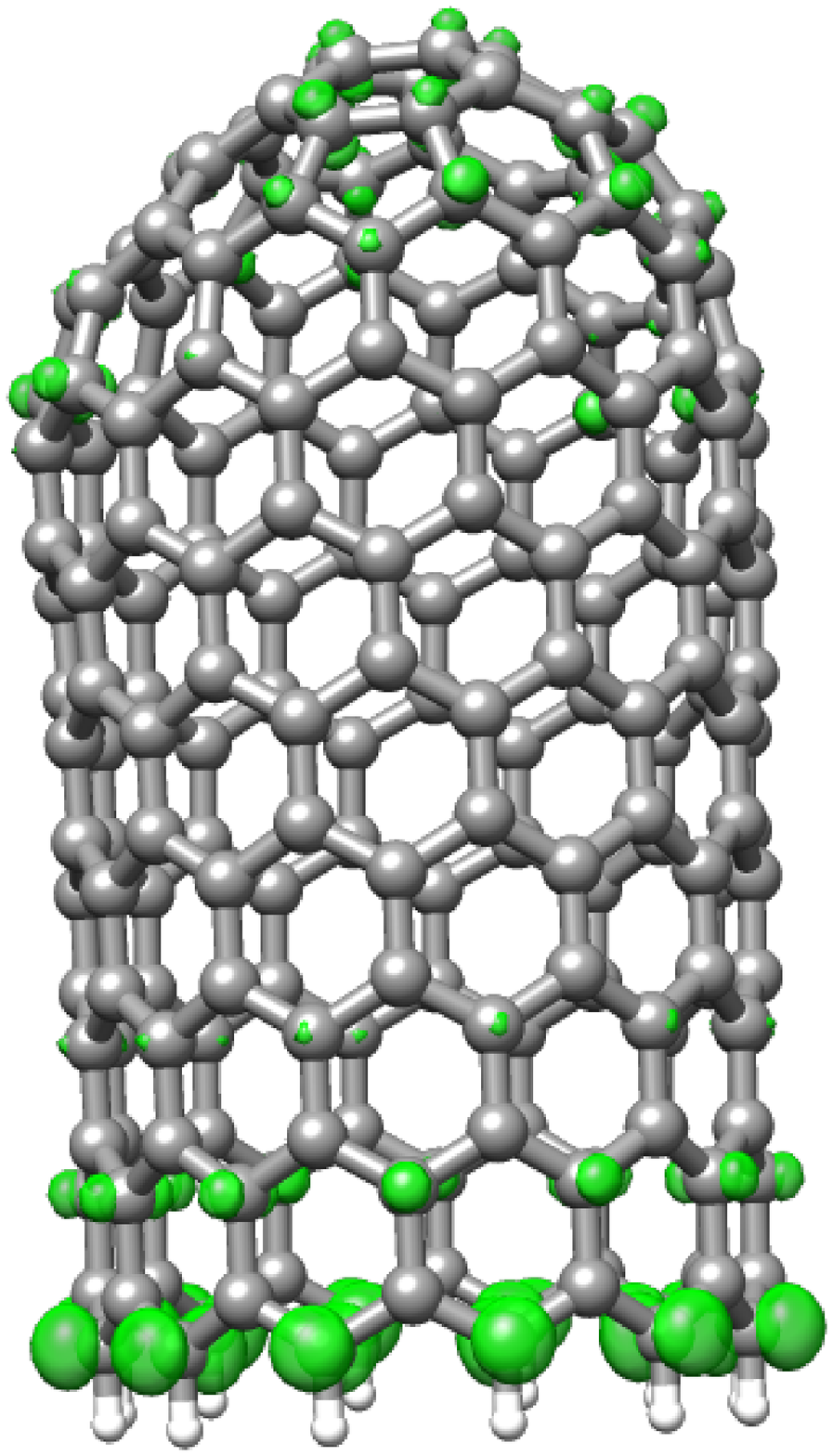}} & \scalebox{0.3}{\includegraphics{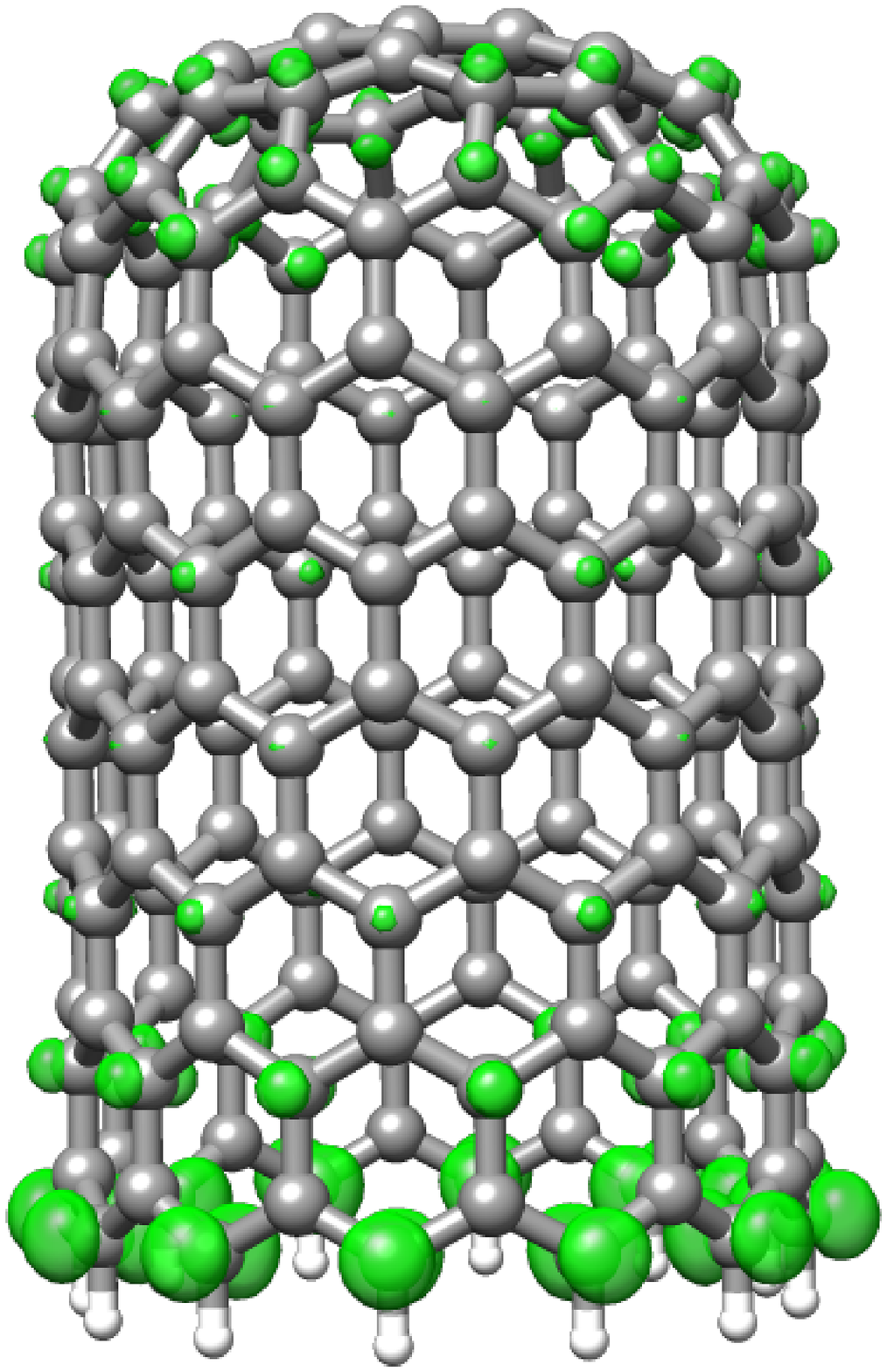}} \\
N$_{\mathrm{FOD}}$ = 8.27 & N$_{\mathrm{FOD}}$ = 8.30 \\
\end{tabular}
\caption{Chemical structures and plots ($\sigma = 0\mathrm{.}005$ e/bohr$^3$) of the FOD density obtained from the FT-DFT method for one-side end-capped SWCNT with pentagonal 
(left) and hexagonal (right) caps base, for both [$6$]CC (top) and [$12$]CC (bottom) systems.
\label{fig:end-capped1}}
\end{center}
\end{figure}

\begin{figure}
\begin{center}
\begin{tabular}{cc}
\scalebox{0.38}{\includegraphics{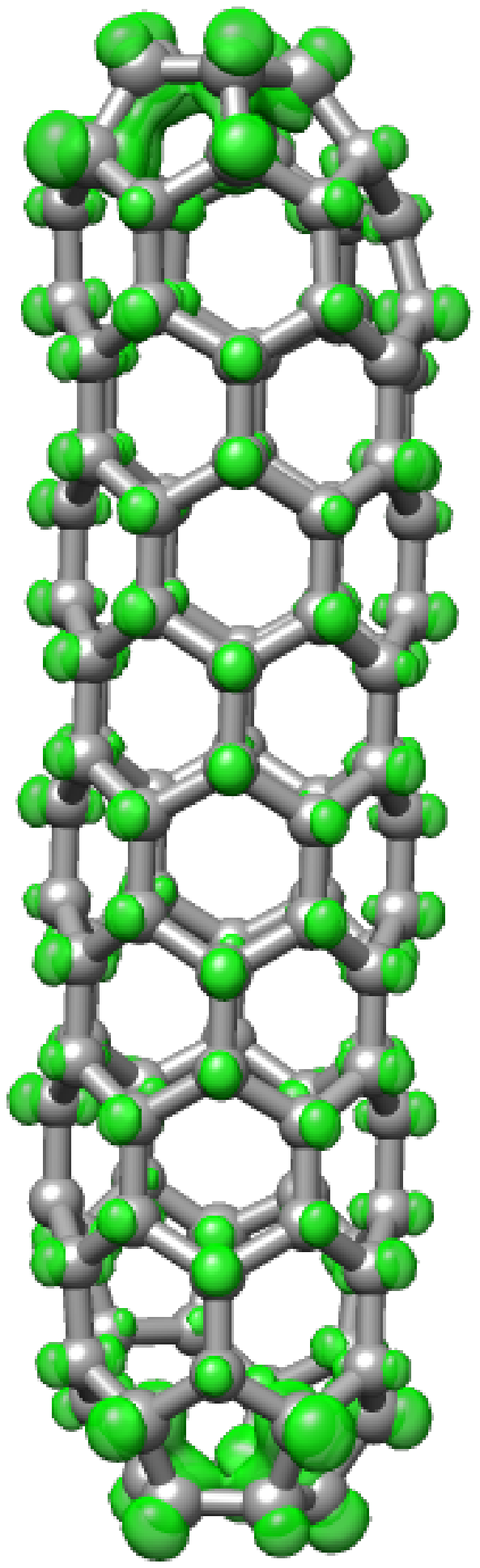}} & \scalebox{0.3}{\includegraphics{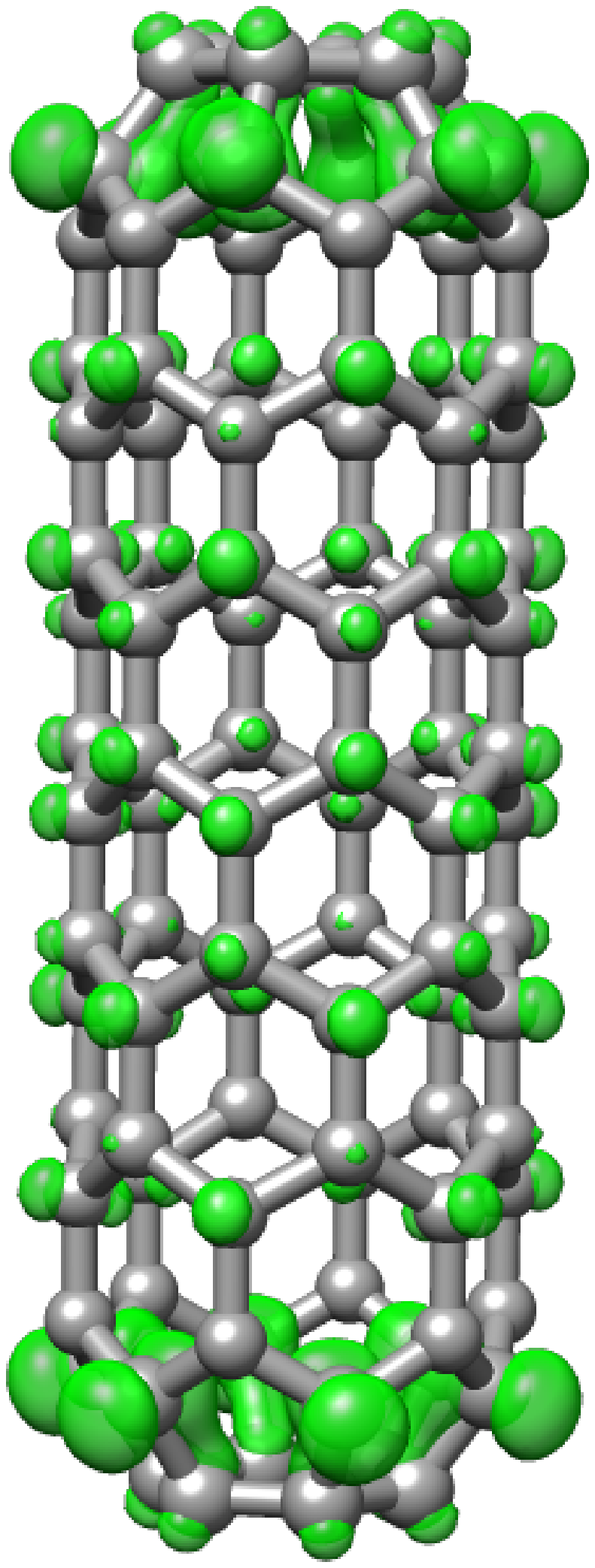}} \\
N$_{\mathrm{FOD}}$ = 7.24 & N$_{\mathrm{FOD}}$ = 6.36 \\ \\
\scalebox{0.35}{\includegraphics{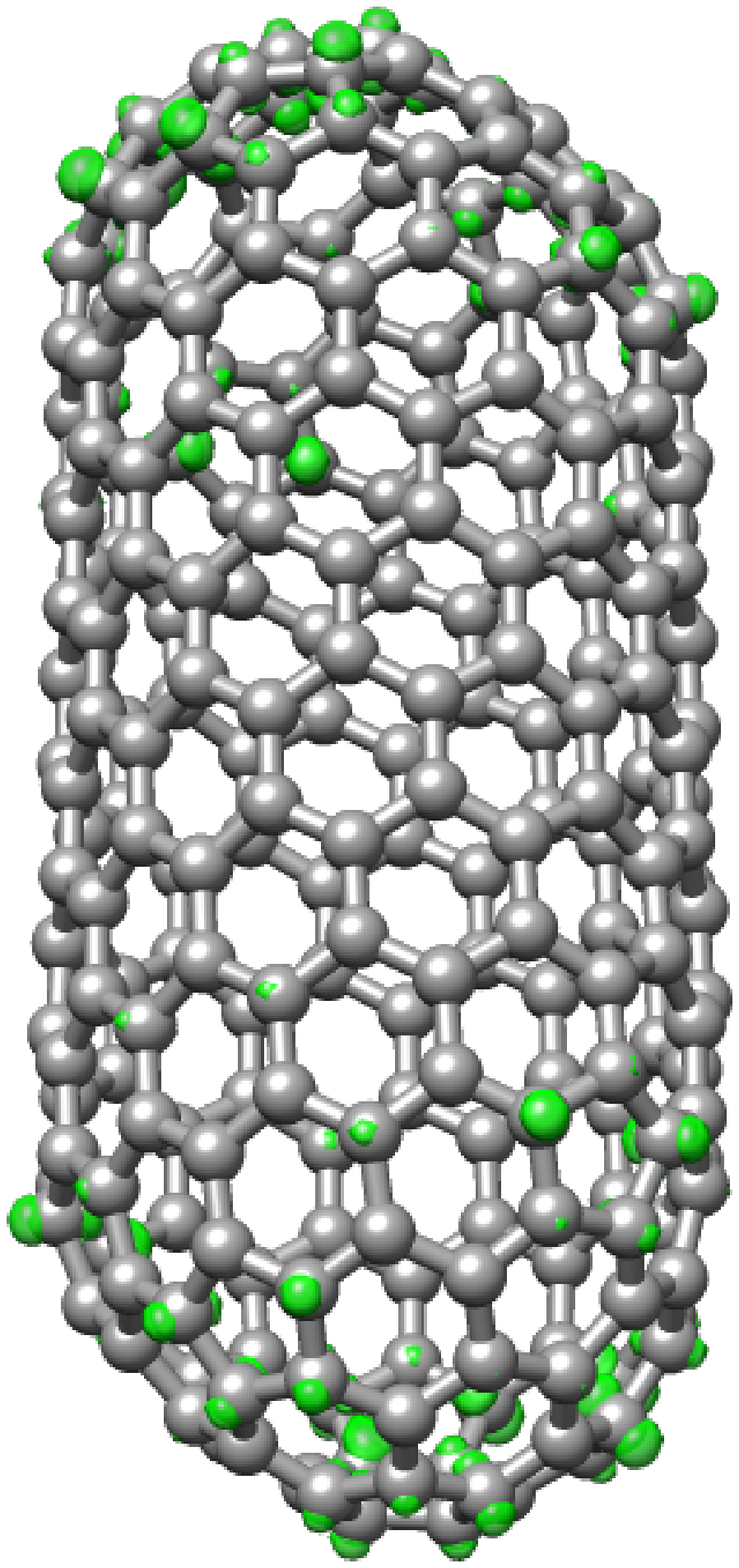}} & \scalebox{0.3}{\includegraphics{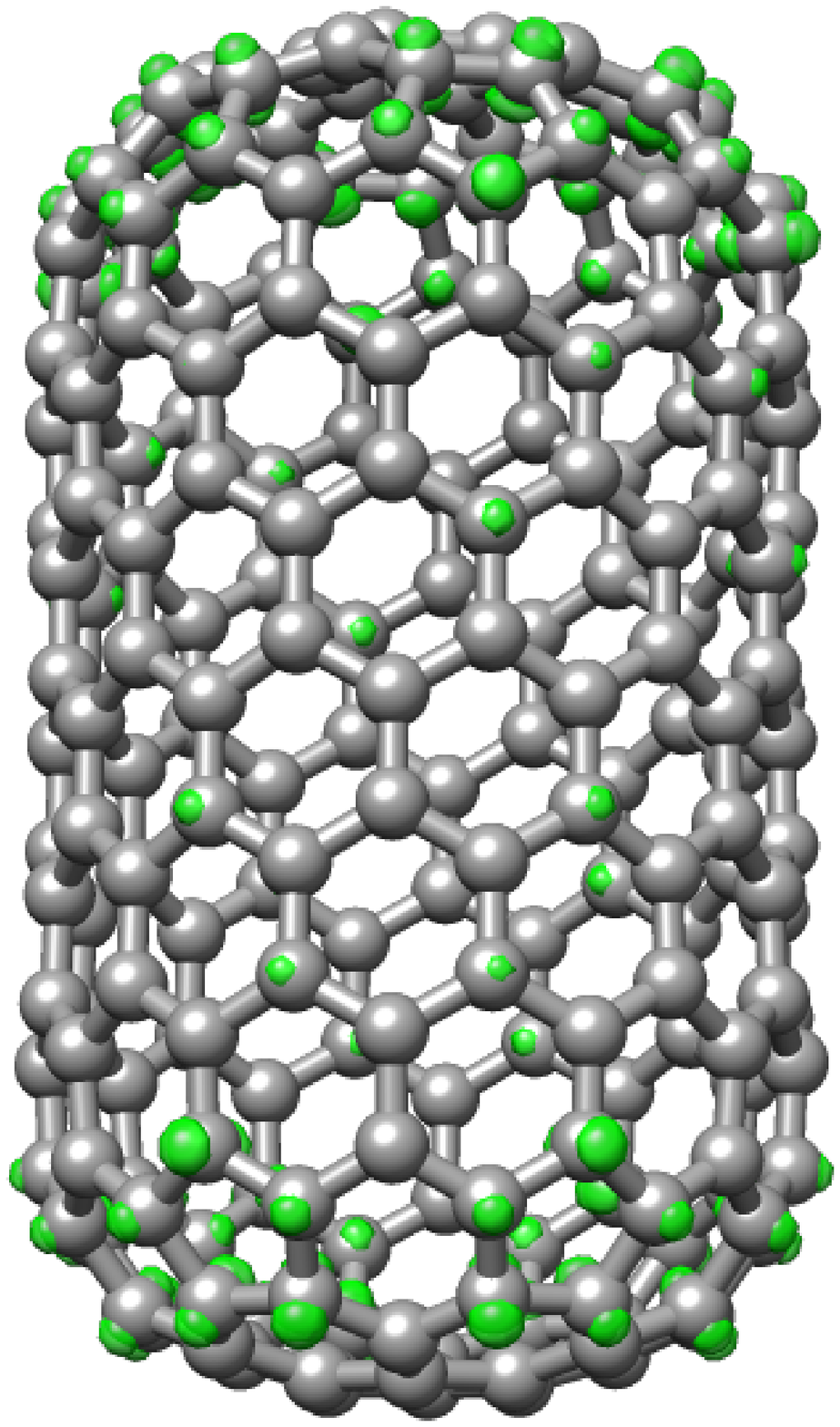}} \\
N$_{\mathrm{FOD}}$ = 8.60 & N$_{\mathrm{FOD}}$ = 7.88 \\
\end{tabular}
\caption{Chemical structures and plots ($\sigma = 0\mathrm{.}005$ e/bohr$^3$) of the FOD density obtained from the FT-DFT method for two-sides end-capped SWCNT with pentagonal
(left) and hexagonal (right) caps base, for both [$6$]CC (top) and [$12$]CC (bottom) systems.
\label{fig:end-capped2}}
\end{center}
\end{figure}

Figure~\ref{fig:caps} displays the $\rho^{\mathrm{FOD}}({\mathbf r})$ distribution and the associated N$_{\mathrm{FOD}}$ values for the isolated caps matching the zigzag edges 
of both [$6$]CC and [$12$]CC, at the FT-TPSS/def2-TZVP//TPSS-D3(BJ)/def2-SVP level. We remark how the pentagon base induces a highly delocalized FOD density and a larger 
N$_{\mathrm{FOD}}$ value than the hexagon base in all cases, in agreement with what one would expect from the pentagon rule establishing that pentagon-pentagon contacts are 
energetically more costly\cite{kobayashi1997}. When these caps are covalently bound to one of the edges of the [$6$]CC and [$12$]CC finite nanotubes ($L=6$), the distribution 
of $\rho^{\mathrm{FOD}}({\mathbf r})$, see Figure~\ref{fig:end-capped1}, resembles that found before for non-capped tubes with a concentration of the FOD density at the free 
edge. Note also the geometric deformation experienced for the tube capped with a pentagon base cap, due to the mismatch between the symmetry of both moieties.
Interestingly, we note that always N$_{\mathrm{FOD}}$ ([$n$]CC@CAP) $<$ N$_{\mathrm{FOD}}$ ([$n$]CC) + N$_{\mathrm{FOD}}$ (CAP), arising from the passivation of edge 
effects which in turn attenuates the (poli-)radical character of the systems. In other words, the nature of the [$n$]CC@CAP is determined primarily by the nature of the tube. When 
two caps are linked to both edges of the finite nanotube, see Figure~\ref{fig:end-capped2}, the attenuation of the (poli-)radical character is specially remarkable for the 
[$12$]CC case, N$_{\mathrm{FOD}}$ ([$12$]CC@CAP$_2$) $<<$ N$_{\mathrm{FOD}}$ ([$12$]CC) + 2 $\cdot$ N$_{\mathrm{FOD}}$ (CAP), and actually N$_{\mathrm{FOD}}$ ([$12$]CC@CAP$_2$) 
$<$ N$_{\mathrm{FOD}}$ ([$12$]CC) despite the larger number of C atoms in [$12$]CC@CAP$_2$, and with values close to those found for the [$6$]CC@CAP$_2$ case. Note also that 
N$_{\mathrm{FOD}}$ ([$n$]CC@CAP) values are always lower when the hexagon base is used to build the cap, for all the cases studied.
\\

\section{Conclusions} 

We have theoretically studied the use of cyclic organic molecules as templates for the controlled growth of carbon nanotubes of finite and well-defined sizes and edges, 
an issue particularly challenging in the case of zig-zag SWCNT due to the (poly-)radical nature of the precursors (i.e. cyclic oligoacenes). Upon a layer-by-layer extension of the 
nanotube size we have demonstrated that zig-zag nanotubes inherit the (poly-)radical character of the constituting moieties, showing a smooth but gradual decrease of 
the singlet-triplet energy difference and an increase of the number of unpaired electrons as a function of the number of layers, as well as accessing to other chemically 
important information not easy to obtain by other techniques. Interestingly, the density arising from the 
fractionally occupied orbitals, independently of the method used for their calculation, shows a strong adherence to the edge in the case of polyradicaloids independently of 
their system size and diameter. On the other hand, armchair nanotubes do not show such a pronounced (poly-)radical character since the constituting unit (i.e. cyclic phenacenes) 
already had an attenuated (poly-)radical character, revealing how the topology of the molecular template is pivotal for the electronic structure of the whole nanotube, and 
actually delocalizing the density arising from the fractionally occupied orbitals along all the structure. We have also investigated the effect of capping the zig-zag finite-size 
nanotube with tailored caps, thus inducing a weak passivation of the (poly-)radical nature disclosed before. Overall, our calculations show the intricate and delicate 
interplay between structural and electronic effects in carbon-based nanoforms, particularly challenging in the case of polyradicaloid molecules.
\\

\section*{Acknowledgements}
A.J.P.J. and J.C.S.G acknowledge the project CTQ2014-55073-P from Spanish Government (MINECO/FEDER) and the project AICO/2018/175 from the Regional Government (GVA/FSE). 
J.F.R. acknowledges the projects MAT2016-78625 from Spanish Government (MINECO/FEDER) and projects No. PTDC/FIS-NAN/4662/2014 and No. PTDC/FIS-NAN/3668/2014 from Portuguese 
Government (Funda\c c\~ ao para a Ci\^ encia e Tecnologia). D.C. is thankful to projects IT588-13 (Eusko Jaurlaritza) and CTQ2016-80955 from the Spanish Government 
(MINECO/FEDER). M.E.S.-S. acknowledges CONACyT-M\'exico for a Ph.D. fellowship (ref. 591700). R.O.C. acknowledges ``Generalitat Valenciana'' and ``Fondo Social Europeo'' 
for a Ph.D. fellowship (ACIF/2018/198).

\section*{Associated content}

The Supporting Information contains in this order: (i) Sketch showing the formation of zero-energy modes when even (left) and odd (right) symmetrical and equivalent parts 
of [$n$]CC are bound together; (ii) Singlet-triplet energy gaps and N$_{\mathrm{FOD}}$ of [$12$]CC with $L = 1-6$ computed at the RAS-SF/6-31G(d) level; (iii) Chemical 
structures and plots ($\sigma = 0\mathrm{.}005$ e/bohr$^3$) of the FOD density as obtained from the FT-DFT method, for the set of [$n$]CC and [$n$]CPH compounds ($L=1-6$); 
(iv) Chemical structures and plots ($\sigma = 0\mathrm{.}002$ e/bohr$^3$) of the FOD density as obtained from the FT-DFT method for the set of [$n$]CPH compounds ($L=1-6$); 
and (v) Chemical structures and plots ($\sigma = 0\mathrm{.}002$ e/bohr$^3$) of the FOD density as obtained from the RAS-SF method, for the set of [$12$]CC compounds ($L=1-6$).
\\

\bibliography{bibliography}

\clearpage
\topmargin0cm

\centering
\begin{sidewaystable}
\begin{threeparttable}
\label{tbl:1}
Table 1: N$_{\mathrm{FOD}}$ and energy magnitudes (in eV) for all systems as a function of their size ($L=1-6$).
\setlength{\extrarowheight}{0.2cm}
\begin{tabular}{llccccccccc}
\hline
 System & Size & N$_{\mathrm{FOD}}$ & $\Delta E_{\mathrm{ST}}$ & VIP & VEA & QEG & $\mu$ & $\omega$ & $\omega^-$ & $\omega^+$ \\
\hline
$[6]$CC & $L=1$ & 1.806 & 0.398 & 5.90 & 1.57 & 4.33 & --2.16 & 3.22 & 5.36 & 1.62 \\
        & $L=2$ & 3.259 & 0.227 & 5.75 & 2.04 & 3.71 & --1.85 & 4.09 & 6.27 & 2.37 \\
        & $L=3$ & 4.097 & 0.171 & 5.65 & 2.33 & 3.32 & --1.66 & 4.79 & 7.00 & 3.00 \\
        & $L=4$ & 4.421 & 0.150 & 5.46 & 2.42 & 3.05 & --1.52 & 5.09 & 7.25 & 3.31 \\
        & $L=5$ & 5.081 & 0.134 & 5.42 & 2.60 & 2.82 & --1.41 & 5.69 & 7.17 & 3.87 \\
        & $L=6$ & 5.767 & 0.122 & 5.38 & 2.75 & 2.63 & --1.32 & 6.27 & 8.46 & 4.40 \\
$[9]$CC & $L=1$ & 3.471 & 0.250 & 5.43 & 1.97 & 3.46 & --1.73 & 3.96 & 6.03 & 2.33 \\
        & $L=2$ & 4.370 & 0.182 & 5.35 & 2.28 & 3.07 & --1.53 & 4.74 & 6.84 & 3.02 \\
        & $L=3$ & 5.027 & 0.149 & 5.27 & 2.48 & 2.79 & --1.40 & 5.39 & 7.50 & 3.62 \\
        & $L=4$ & 5.781 & 0.129 & 5.21 & 2.64 & 2.57 & --1.29 & 5.99 & 8.11 & 4.19 \\
        & $L=5$ & 6.443 & 0.116 & 5.16 & 2.77 & 2.40 & --1.20 & 6.56 & 8.69 & 4.72 \\
        & $L=6$ & 7.036 & 0.107 & 5.12 & 2.87 & 2.25 & --1.13 & 7.09 & 9.23 & 5.24 \\
\hline
\end{tabular}
\end{threeparttable}
\end{sidewaystable}

\clearpage
\topmargin0cm

\centering
\begin{sidewaystable}
\begin{threeparttable}
\label{tbl:1}
Table 1 (cont.): N$_{\mathrm{FOD}}$ and energy magnitudes (in eV) for all systems as a function of their size ($L=1-6$).
\setlength{\extrarowheight}{0.2cm}
\begin{tabular}{llccccccccc}
\hline
 System & Size & N$_{\mathrm{FOD}}$ & $\Delta E_{\mathrm{ST}}$ & VIP & VEA & QEG & $\mu$ & $\omega$ & $\omega^-$ & $\omega^+$ \\
\hline
$[12]$CC  & $L=1$ & 3.212 & 0.227 & 5.24 & 2.28 & 2.95 & --1.48 & 4.78 & 6.85 & 3.09 \\
          & $L=2$ & 5.653 & 0.141 & 5.14 & 2.52 & 2.61 & --1.31 & 5.61 & 7.69 & 3.86 \\
          & $L=3$ & 6.766 & 0.116 & 5.08 & 2.69 & 2.40 & --1.20 & 6.30 & 8.39 & 4.51 \\
          & $L=4$ & 7.588 & 0.102 & 5.04 & 2.81 & 2.23 & --1.11 & 6.93 & 9.03 & 5.10 \\
          & $L=5$ & 8.261 & 0.093 & 5.00 & 2.91 & 2.09 & --1.04 & 7.51 & 9.62 & 5.66 \\
          & $L=6$ & 8.863 & 0.087 & 4.97 & 3.00 & 1.97 & --0.98 & 8.08 & 10.2 & 6.21 \\
$[12]$CPH & $L=1$ & 0.768 & 0.856 & 5.54 & 1.87 & 3.67 & --1.84 & 3.74 & 5.82 & 2.11 \\
          & $L=2$ & 1.842 & 0.495 & 5.29 & 2.26 & 3.03 & --1.52 & 4.71 & 6.79 & 3.01 \\
          & $L=3$ & 2.475 & 0.363 & 5.16 & 2.48 & 2.69 & --1.34 & 5.44 & 7.52 & 3.69 \\
          & $L=4$ & 2.786 & 0.310 & 5.11 & 2.65 & 2.46 & --1.23 & 6.11 & 8.20 & 4.32 \\
          & $L=5$ & 4.137 & 0.252 & 5.04 & 2.77 & 2.26 & --1.13 & 6.74 & 8.83 & 4.93 \\
          & $L=6$ & 4.183 & 0.221 & 5.00 & 2.89 & 2.11 & --1.06 & 7.36 & 9.46 & 5.52 \\
\hline  
\end{tabular}
\end{threeparttable}
\end{sidewaystable}

\clearpage
\topmargin0cm

\clearpage
\topmargin0cm

\clearpage
\topmargin0cm

\clearpage
\topmargin0cm

\clearpage
\topmargin0cm

\clearpage
\topmargin0cm

\clearpage
\topmargin0cm

\clearpage
\topmargin0cm

\clearpage
\topmargin0cm

\clearpage
\topmargin0cm

\end{document}